\documentclass[12pt,fleqn]{article}
\usepackage[cp866]{inputenc}
\usepackage[english]{babel}
\usepackage{makeidx}
\usepackage{latexsym,amsfonts,amssymb,amsmath,longtable,amsthm}
\input amssym.def
\usepackage{latexsym}
\usepackage{graphicx}
\usepackage[all]{xy}
\usepackage{amsfonts}
\usepackage{amsmath}
\usepackage{mathrsfs}
\usepackage{color}
\usepackage{ulem}

\def\w{{\mathbf{w}}}
\def\u{{\mathbf{u}}}
\def\hw{{\widehat{\mathbf{w}}}}
\def\hu{{\widehat{\mathbf{u}}}}
\def\hp{{\widehat{p}}}
\def\U{{\mathbf{U}}}
\def\D{{\mathcal{D}}}

\def\Ha{{\mathfrak{H}}}

\def\F{{\mathcal{F}}}

\def\F{{\mathcal{F}}}

\def\bfeta{\boldsymbol{\eta}}

\def\int{\intop}

\newcommand{\wotwo}{\mathop{\mathaccent23{W}^{1,2}}\nolimits}

\newcommand{\woq}{\mathop{\mathaccent23{W}^{k,q}}\nolimits}

\def\div{\hbox{\rm div}\,}
\def\curl{\hbox{\rm curl}}

\newtheorem{theo}{\bf Theorem}
\newtheorem{lem}{\bf Lemma}

\newtheorem{lemr}{\bf Remark}

\newcommand{\pr}{{\bf Proof. }}

\def\Xint#1{\mathchoice
   {\XXint\displaystyle\textstyle{#1}}%
   {\XXint\textstyle\scriptstyle{#1}}%
   {\XXint\scriptstyle\scriptscriptstyle{#1}}%
   {\XXint\scriptscriptstyle\scriptscriptstyle{#1}}%
   \!\int}
\def\XXint#1#2#3{{\setbox0=\hbox{$#1{#2#3}{\int}$}
     \vcenter{\hbox{$#2#3$}}\kern-.5\wd0}}

\def\dashint{\Xint-}

\newcommand{\esssup}{\mathop{\rm ess\,sup}}
\newcommand{\R}{{\mathbb R}}
\newcommand{\const}{{\rm const}}
\newcommand{\dist}{\mathop{\rm dist}}
\newcommand{\arctg}{\mathop{\rm arctg}}
\newcommand{\Int}{\mathop{\rm Int}}
\newcommand{\Cl}{\mathop{\rm Cl}}
\newcommand{\meas}{\mathop{\rm meas}}

\newcommand{\ve}{{\mathbf v}}
\newcommand{\diam}{\mathop{\rm diam}}
\renewcommand{\div}{\mathop{\rm div}}

\newcommand{\loc}{{\rm loc}}

\begin{document}
\title{The existence theorem for  steady
Navier--Stokes equations  in the axially symmetric case
\footnote{
domains,  Stokes system, stationary Navier Stokes equations,
boundary--value problem.}}
\author{Mikhail Korobkov\footnote{Sobolev Institute of Mathematics,
Koptyuga pr. 4, and Novosibirsk State University, Pirogova Str. 2,
630090 Novosibirsk, Russia; korob@math.nsc.ru}, Konstantin
Pileckas\footnote{Faculty of Mathematics and Informatics, Vilnius
University, Naugarduko Str., 24, Vilnius, 03225  Lithuania;
pileckas@ktl.mii.lt},  and Remigio Russo\footnote{ Dipartimento di
Matematica, Seconda Universit\`a di Napoli,  via Vivaldi 43, 81100
Caserta, Italy; remigio.russo@unina2.it}$\;$}

\maketitle

\begin{abstract} We study the nonhomogeneous boundary value problem
for Navier--Stokes equations of steady motion of a viscous
incompressible fluid in  a three--dimensional   bounded domain
with multiply connected boundary. We prove that this problem has a
solution in some axially symmetric cases, in particular, when all
components of the boundary intersect the axis of symmetry.
\end{abstract}

\bigskip
\setcounter{section}{0}
\section{Introduction}
Let $\Omega$ be a bounded domain in $\R^3$ with with multiply
connected Lipschitz boundary $\partial\Omega$ consisting of $N+1$
disjoint components $\Gamma_j$:
$\partial\Omega=\Gamma_0\cup\ldots\cup\Gamma_N$, and
$\Gamma_i\cap\Gamma_j=\emptyset,\, i\neq j$. Consider in $\Omega$
the stationary  Navier--Stokes system  with nonhomogeneous
boundary conditions
\begin{equation}
\label{NS}
 \left\{\begin{array}{rcl}
-\nu \Delta{\bf u}+\big({\bf u}\cdot \nabla\big){\bf u} +\nabla p
&
= & {0}\qquad \hbox{\rm in }\;\;\Omega,\\[4pt]
\div\,{\bf u} &  = & 0  \qquad \hbox{\rm in }\;\;\Omega,
\\[4pt]
 {\bf u} &  = & {\bf a}
 \qquad \hbox{\rm on }\;\;\partial\Omega.
\end{array}\right.
\end{equation}

The continuity equation $(\ref{NS}_2)$ implies the necessary
compatibility condition for the solvability of problem (\ref{NS}):
\begin{equation}\label{flux}
\int\limits_{\partial\Omega}{\bf a}\cdot{\bf
n}\,dS=\sum\limits_{j=0}^N\int\limits_{\Gamma_j}{\bf a}\cdot{\bf
n}\,dS=\sum\limits_{j=0}^N{\mathcal F}_i=0,
\end{equation} where ${\bf n}$ is a unit vector
of the outward (with respect to $\Omega$) normal to
$\partial\Omega$ and ${\mathcal F}_j=\int\limits_{\Gamma_j}{\bf
a}\cdot{\bf n}\,dS$.

Starting from the famous  paper of J. Leray \cite{Leray} published
in 1933, problem (\ref{NS}) was a subject of investigation in many
papers (see, e.g., \cite{Amick}, \cite{BOPI},
\cite{Finn}--\cite{KaPi1}, \cite{Kozono}--\cite{Lad},
\cite{Morimoto}--\cite{VorJud}, etc.). However, for a long time
the existence of a weak solution ${\bf u}\in W^{1,2}(\Omega)$ to
problem (\ref{NS}) was proved only under the condition
\begin{equation}\label{flux0}
{\F}_j=\int\limits_{\Gamma_j}{\bf a}\cdot{\bf n}\,dS=0,\qquad
j=1,2,\ldots,N,
\end{equation}
or for sufficiently small fluxes (see  \cite{Leray},
\cite{Lad1}--\cite{Lad}, \cite{Fu}, \cite{VorJud}, \cite{Kozono},
etc.). Condition (\ref{flux0}) requires the net flux ${\F}_j$ of
the boundary value ${\bf a}$ to be zero separately across each
component $\Gamma_j$ of the boundary $\partial\Omega$, while the
compatibility condition (\ref{flux}) means only that the total
flux is zero. Thus, (\ref{flux0}) is stronger than (\ref{flux})
(condition (\ref{flux0}) does not allow the presence of sinks and
sources).

For a detailed survey of previous results one can see the recent
papers~\cite{kpr} or \cite{Pukhnachev}--\cite{Pukhnachev1}. In
particular, in the last papers V.V.~Pukhnachev  has established
the existence of a~solution to problem (\ref{NS}) in the
three--dimensional case when the domain $\Omega$ and the boundary
value ${\bf a}$ have an axis of symmetry and a plane of symmetry
which is perpendicular to this axis, moreover, this plane
intersects each component of the boundary.

In this paper we study the problem in the axial symmetric case.
Let $O_{x_1},O_{x_2},O_{x_3}$ be coordinate axis in $\R^3$  and
$\theta=\arctg(x_2/x_1)$, $r=(x_1^2+x_2^2)^{1/2}$, $z=x_3$ be
cylindrical coordinates. Denote by $v_\theta,v_r,v_z$ the
projections of the vector $\bf v$ on the axes $\theta,r,z$.

A function $f$ is said to be {\it axially symmetric} if it does
not depend on~$\theta$. A vector-valued function ${\bf
h}=(h_\theta,h_r,h_z)$ is called {\it axially symmetric} if
$h_\theta$, $h_r$ and $h_z$ do not depend on~$\theta$. A
vector-valued function ${\bf h}=(h_\theta,h_r,h_z)$ is called {\it
axially symmetric without rotation} if $h_\theta=0$ while $h_r$
and $h_z$ do not depend on~$\theta$.

We will use the following symmetry assumptions.

(SO) $\Omega\subset \R^3$ is a bounded domain with Lipschitz
boundary and $O_{x_3}$ is the axis of symmetry of the domain
$\Omega$.

(AS) The assumptions {\rm (SO)} are fulfilled and the boundary
value ${\bf a}\in W^{1/2,2}(\partial\Omega)$ is axially symmetric.

(ASwR) The assumptions {\rm (SO)} are fulfilled and the boundary
value ${\bf a}\in W^{1/2,2}(\partial\Omega)$ is axially symmetric
without rotation.

Denote by $\Omega_j$ the bounded simply connected domain with
$\partial\Omega_j=\Gamma_j$, $j=0,\dots,N$. Let $\Omega_0$ be the
largest domain, i.e.,
$$\Omega=\Omega_0\setminus\bigl(\cup_{j=1}^N\bar\Omega_j\bigr).$$
Here and henceforth we denote by $\bar A$ the closure of the
set~$A$.

Let
$$\Gamma_j\cap O_{x_3}\ne\emptyset,\quad j=0,\dots,M,$$
$$\Gamma_j\cap O_{x_3}=\emptyset,\quad j=M+1,\dots,N.$$
We shall prove the existence theorem if one of the following two
additional conditions is fulfilled:
\begin{equation}\label{flux^1}
M=N-1, \qquad \F_N\ge0,
\end{equation}
or
\begin{equation}\label{flux^2}
|\F_j|<\delta,\quad j=M+1, \ldots, N,
\end{equation}
where $\delta=\delta(\nu,\Omega)$ is sufficiently small
($\delta(\nu,\Omega)$  is specified below  in Section~\ref{poet}).
In particular, (\ref{flux^2}) includes the case~$N=M$, i.e, when
each component of the boundary intersects the axis of symmetry.
Notice that in (\ref{flux^1}), (\ref{flux^2}) the fluxes $\F_j,
j=1,\ldots, M$, could be arbitrary large.

\pagebreak
\begin{figure}[htp]
\begin{center}
{(a) M=N=2}\includegraphics[height=4.cm]{u1.eps} {(b) M=2,
N=3}\includegraphics[height=4.cm]{u2.eps} {(c)~M=1,
N=3}\includegraphics[height=4.cm]{u3.eps}
\end{center}
\caption{Domain $\Omega$}
  \label{fig:edge}
\end{figure}

On Fig.1 we show several possible domains $\Omega$. In the case
(a) all fluxes $\F_0, \F_1$ and $\F_2$ are arbitrary; in the case
(b) fluxes $\F_0, \F_1, \F_2$ are arbitrary, while the flux $\F_3$
has to be nonnegative, but there are no restriction on its size;
in the case (c) fluxes $\F_0, \F_1$ are arbitrary,
while $\F_2$ and $\F_3$ has to be "sufficiently small".   \\

The main result of the paper reads as follows.

\begin{theo} \label{kmpTh4.2} {\sl
Let the conditions {\rm (AS)},  (\ref{flux}) be fulfilled. Suppose
 that one of the conditions~(\ref{flux^1})
or~(\ref{flux^2}) holds. Then the problem $(\ref{NS})$ admits at
least one weak axially symmetric solution ${\bf u}\in
W^{1,2}(\Omega)$.

If, in addition, the conditions {\rm (ASwR)} are fulfilled, then
the problem $(\ref{NS})$ admits at least one weak axially
symmetric solution without rotation. }
\end{theo}

(For the definition of a weak solution, see Section~2.1.) The
analogous results for the plane case were established
in~\cite{kpr}.

The proof of Theorem~\ref{kmpTh4.2} uses the Bernoulli law for a
weak solution of the Euler equations and the one-side maximum
principle for the total head pressure corresponding to this
solution (see Section~\ref{eueq}).  These results were obtained
in~\cite{korob1} for plane case (see ~\cite{kpr} for more detailed
proofs). The proof of the Bernoulli law for solutions from Sobolev
spaces is based on recent results obtained in~\cite{korob} (see
also Section~\ref{SMS}).

The short version of this paper was published in \cite{kpr_a_crm}.
The preprint version of this paper see in \cite{kpr_a_arx}.

\section{Notations and preliminary results}
\setcounter{theo}{0} \setcounter{lem}{0}
\setcounter{lemr}{0}\setcounter{equation}{0}

By {\it a domain} we mean an open connected set. Let
$\Omega\subset\R^3$ be a bounded domain with Lipschitz boundary
$\partial\Omega$. We use standard notations for function spaces:
$C^k(\bar\Omega)$, $C^k(\partial\Omega)$, $W^{k,q}(\Omega)$,
$\woq(\Omega)$, $W^{\alpha,q}(\partial\Omega)$, where
$\alpha\in(0,1), k\in{\mathbb N}_0, q\in[1,+\infty]$. In our
notation we do not distinguish function spaces for scalar and
vector valued functions; it is clear from the context whether we
use scalar or vector (or tensor) valued function spaces.
$H(\Omega)$ is subspace of all solenoidal  vector fields
($\div{\bf u}=0$) from $\wotwo(\Omega)$ with the norm $\|{\bf
u}\|_{H(\Omega)}=\|\nabla{\bf u}\|_{L^2(\Omega)}$.  Note that for
functions ${\bf u}\in H(\Omega)$ the norm $\|\cdot\|_{H(\Omega)}$
is equivalent to $\|\cdot\|_{W^{1,2}(\Omega)}$.

Working with Sobolev functions we always assume that the "best
representatives" are chosen. If $w\in L^1_{\loc}(\Omega)$, then
the best representative $w^*$ is defined by
\begin{displaymath} w^*(x)=\left\{\begin{array}{rcl}\lim\limits_{r\to
0} \dashint_{B_r(x)}{
w}(z)dz, & {\rm \;if\; the\; finite\; limit\; exists;} \\[4pt]
 0 \qquad\qquad\quad & \; {\rm otherwise },
\end{array}\right.
\end{displaymath}
where $\dashint_{B_r(x)}{ w}(z)dz=\frac{1}{\meas(
B_r(x))}\int_{B_r(x)}{ w}(z)dz$, $B_r(x)=\{y: |y-x|<r\}$ is a ball
of radius $r$ centered at $x$.

Further (see Theorem~\ref{kmpTh2.1}) we will discuss some
properties of the best representatives of Sobolev functions.

\subsection{Some facts about solenoidal functions}

The next lemmas concern the existence of a solenoidal extensions
of boundary values  and the integral representation of the bounded
linear functionals vanishing on solenoidal functions.

\begin{lem} [see Corollary~2.3 in \cite{LadSol1}]
\label{kmpLem14.1}  {\sl Let $\Omega\subset\R^3 $ be a bounded
domain with Lipschitz boundary. If ${\bf a}\in
W^{1/2,2}(\partial\Omega)$ and the equality~(\ref{flux}) is
fulfilled, then there exists a solenoidal extension ${\bf A}\in
W^{1,2}(\Omega)$ of ${\bf a}$ such that
\begin{equation}
\label{is2} \|{\bf A}\|_{W^{1,2}(\Omega)}\leq c \|{\bf
a}\|_{W^{1/2,2}(\partial\Omega)}.
\end{equation}
}\end{lem}

From this Lemma we can deduce some assertions for the symmetric
case.

\begin{lem}
\label{kmpLem14.1'}  {\sl Let the conditions {\rm (AS)},
(\ref{flux}) be fulfilled. Then there exists an axially symmetric
solenoidal extension ${\bf A}\in W^{1,2}(\Omega)$ of ${\bf a}$
such that the estimate~(\ref{is2}) holds. }\end{lem} \pr Let ${\bf
A_0}\in W^{1,2}(\Omega)$ be solenoidal extension of ${\bf a}$ from
Lemma~\ref{kmpLem14.1}. Put
$${\bf A}_i(\theta,r,z)=\frac1{i!}\sum\limits_{j=0}^{i!}{\bf A_0}\bigl(\theta+\frac{2\pi
j}{i!},r,z\bigr).$$ Clearly, each ${\bf A}_i$ is also solenoidal
extension of ${\bf a}$ and the estimate~(\ref{is2}) holds for
${\bf A}_i$ with the same~$c$ (not depending on~$i$). By
construction
\begin{equation}
\label{is2.1}{\bf A}_i(\theta+\frac{2\pi j}m,r,z)={\bf
A}_i(\theta,r,z)\quad\mbox{for all } m=1,\dots,i.
\end{equation}
Take a weakly convergence sequence ${\bf A}_{i_k}\rightharpoonup
{\bf A}$ in $W^{1,2}(\Omega)$. Then by construction $\div{\bf
A}=0$, \,${\bf A}|_{\partial\Omega}={\bf a}$, and the
estimate~(\ref{is2}) holds. From~(\ref{is2.1}) it follows that
${\bf A}(\theta+\frac{2\pi j}m,r,z)={\bf A}(\theta,r,z)$ for all
$m,j$. Hence $\bf A$ is axially symmetric.  $\qed$

\begin{lem}
\label{kmpLem14.1''}{\sl  Let the conditions {\rm (ASwR)},
(\ref{flux}) be fulfilled. Then there exists a~solenoidal
extension ${\bf A}\in W^{1,2}(\Omega)$ of ${\bf a}$ such that
${\bf A}$ is axially symmetric without rotation and the
estimate~(\ref{is2}) holds.}
\end{lem}
\pr Let ${\bf \tilde A}=(\tilde A_\theta,\tilde A_r,\tilde A_z)
\in W^{1,2}(\Omega)$ be solenoidal extension of ${\bf a}$ from the
previous Lemma~\ref{kmpLem14.1'}. Then by classical formula
\begin{equation}
\label{is2.2}
\begin{array}{rcl}
\operatorname{div}\mathbf{\tilde A}( \theta, r,z) =
\frac{1}{r}\frac{\partial}{\partial \theta}(\tilde
A_\theta)+\frac{1}{r} \frac{\partial}{\partial r}(\tilde A_r r) +
\frac{\partial}{\partial z}(\tilde A_z)\\[4pt]
=\frac{1}{r} \frac{\partial}{\partial r}(\tilde A_r r) +
\frac{\partial}{\partial z}(\tilde A_z)=0.
\end{array}
\end{equation}
Here $\frac{\partial \tilde A_\theta}{\partial \theta}=0$ because
of axial symmetry. Define the vector field ${\bf
A}=(A_\theta,A_r,A_z)$ by the formulas
$$A_\theta=0,\quad A_r=\tilde A_r,\quad A_z=\tilde A_z.$$
Then by construction  ${\bf A}$ is axially symmetric without
rotation, ${\bf A}|_{\partial\Omega}={\bf a}$, and the
estimate~(\ref{is2}) holds. From~(\ref{is2.2}) it follows that
$\div {\bf A}=0$. $\qed$

\begin{lem} [see \cite{SolSca}]
\label{kmpLem14.2}  {\sl  Let  $\Omega\subset\R^3 $ be a bounded
domain with Lipschitz boundary and  $R(\bfeta)$ be a continuous
linear functional defined on $\wotwo(\Omega)$. If
\begin{displaymath}
R(\bfeta)=0\qquad\forall\;\;\bfeta\in H(\Omega),
\end{displaymath}
then there exists a unique function $p\in L^2(\Omega)$ with
$\int\limits_\Omega p(x)\,dx=0$ such that
\begin{displaymath}
R(\bfeta)= \int\limits_\Omega p\,{\rm
div}\,\bfeta\,dx\qquad\forall\;\;\bfeta\in \wotwo(\Omega).
\end{displaymath}
Moreover, $\|p\|_{L^2(\Omega)}$ is equivalent to
$\|R\|_{(\wotwo(\Omega))^*}$. }\end{lem}

\begin{lem}
\label{kmpLem14.2'}{\sl If, in addition to conditions of  Lemma
\ref{kmpLem14.2}, the domain $\Omega$ satisfies the
assumption~{\rm (SO)} and $R(\bfeta)\equiv R(\bfeta_{\theta_0})$
for all $\bfeta\in H(\Omega)$, $\theta_0\in[0,2\pi]$, where
$\bfeta_{\theta_0}(\theta,r,z):=\bfeta(\theta+\theta_0,r,z)$, then
the function $p$ is axially symmetric. }\end{lem} \pr Take the
function $p$ from the assertion of Lemma~\ref{kmpLem14.2}. For
$\theta_0\in[0,2\pi]$ define the function $p_{\theta_0}$ by the
formula $p_{\theta_0}(\theta,r,z):=p(\theta-\theta_0,r,z)$. By
construction,
\begin{displaymath}
\int\limits_\Omega p\,{\rm
div}\,\bfeta\,dx=R(\bfeta)=R(\bfeta_{\theta_0})=\int\limits_\Omega
p\,{\rm div}\,\bfeta_{\theta_0}\,dx
\end{displaymath}
\begin{displaymath}
= \int\limits_\Omega p_{\theta_0}{\rm div}\,\bfeta\,dx
\qquad\forall\;\;\bfeta\in \wotwo(\Omega).
\end{displaymath}
Since $p$ is unique, we obtain the identity $p(x)\equiv
p_{\theta_0}(x)$. $\qed$

\begin{lem} [see \cite{Lad}]
\label{kmpLem14.3} {\sl Let $\Omega\subset\R^3 $ be a bounded
domain with Lipschitz boundary and  let ${\bf A}\in
W^{1,2}(\Omega)$ be divergence free. Then there exists a unique
weak solution ${\bf U}\in W^{1,2}(\Omega)$ of  the Stokes problem
satisfying the boundary condition ${\bf U}|_{\partial\Omega}={\bf
A}|_{\partial\Omega}$, i.e., ${\bf U}-{\bf A}\in H(\Omega)$ and
\begin{equation}\label{4.1}
\int\limits_\Omega\nabla{\bf U}\cdot\nabla\bfeta\,dx= 0
\quad\forall\;\bfeta\in H(\Omega).
\end{equation}
Moreover,
\begin{equation}\label{4.2}
\|{\bf U}\|_{W^{1,2}(\Omega)}\leq c\|{\bf A}\|_{W^{1,2}(\Omega)}.
\end{equation}
}\end{lem}

\begin{lem}
\label{kmpLem14.3'} {\sl  If, in addition to conditions of Lemma
\ref{kmpLem14.3}, the domain $\Omega$ satisfies the
assumptions~{\rm (SO)} and also $\bf A$ is axially symmetric, then
$\bf U$ is axially symmetric too.}
\end{lem}
\pr Let $\bf U$ be a solution of the Stokes problem from
Lemma~\ref{kmpLem14.3}. For $\theta_0\in[0,2\pi]$ define the
function ${\bf U}_{\theta_0}$ by the formula ${\bf
U}_{\theta_0}(\theta,r,z):={\bf U}(\theta-\theta_0,r,z)$. By
construction, ${\bf U}_{\theta_0}-{\bf A}\in H(\Omega)$. Moreover,
\begin{displaymath}
\int\limits_\Omega\nabla{\bf U}_{\theta_0}\cdot\nabla\bfeta\,dx=
\int\limits_\Omega\nabla{\bf U}\cdot\nabla\bfeta_{\theta_0}\,dx
 =0
\quad\ \forall\;\bfeta\in H(\Omega),
\end{displaymath}
where
$\bfeta_{\theta_0}(\theta,r,z):=\bfeta(\theta+\theta_0,r,z)$.
Because of the uniqueness, we obtain the identity ${\bf
U}(x)\equiv {\bf U}_{\theta_0}(x)$. $\qed$

\begin{lem}
\label{kmpLem14.3''}{\sl  If, in addition to conditions of
Lemma~\ref{kmpLem14.3}, the vector field $\bf A$ is axially
symmetric without rotation, then $\bf U$ is axially symmetric
without rotation too.}
\end{lem}
\pr Take the function ${\bf U}=(U_\theta,U_r,U_z)$ from the
assertion of Lemma~\ref{kmpLem14.3} and define
$\bfeta=(\eta_\theta,\eta_r,\eta_z)$ by the formulas
\begin{displaymath}
\eta_\theta\equiv U_\theta, \quad\eta_r=\eta_z\equiv0.
\end{displaymath}
Then from  Lemma~\ref{kmpLem14.3'} it follows the inclusion
$\bfeta\in H(\Omega)$ (see also the formula~(\ref{is2.2})).
Consequently, from~(\ref{4.1}) we obtain
\begin{equation}\label{4.3ax}
\int\limits_\Omega\nabla{\bf U}\cdot\nabla\bfeta\,dx= 0.
\end{equation}
But by the direct calculation
\begin{equation}\label{4.4ax}
\nabla{\bf U}\cdot\nabla\bfeta\equiv
\biggl(\frac{U_\theta}r\biggr)^2+\biggl(\frac{\partial
U_\theta}{\partial r}\biggr)^2+\biggl(\frac{\partial
U_\theta}{\partial z}\biggr)^2.
\end{equation}
Formulas (\ref{4.3ax}), (\ref{4.4ax}) imply the required
equality~$U_\theta\equiv0$. $\qed$
\\

For  a function ${\bf f}\in L^{q}(\Omega), 1\leq q\leq 6/5$,
consider a continuous linear functional $H(\Omega)\ni
\bfeta\mapsto \int\limits_\Omega{\bf f}\cdot\bfeta\,dx$. Because
of Riesz representation theorem there exists a unique function
${\bf g}\in H(\Omega)$ such that
$$
\int\limits_\Omega{\bf
f}\cdot\bfeta\,dx=\int\limits_\Omega\nabla\bfeta\cdot\nabla{\bf
g}\,dx=\langle{\bf g},\bfeta\rangle_{H(\Omega)}
\quad\forall\bfeta\in H(\Omega).
$$
Denote ${\bf g}=T_0{\bf f}$. Evidently, $T_0$ is a continuous
linear operator from $L^q(\Omega)$ to $H(\Omega)$.

Denote by $L^q_{AS}(\Omega)$ the space of all axially symmetric
vector-function from $L^q(\Omega)$. Analogously define the spaces
$L^q_{ASwR}(\Omega)$,  $H_{AS}(\Omega)$, $H_{ASwR}(\Omega)$,
$W^{1,2}_{AS}(\Omega)$, $W^{1,2}_{ASwR}(\Omega)$, etc.

\begin{lem}
\label{kmpLem14.4} {\sl The operator $T_0:L^{3/2}(\Omega)\to
H(\Omega)$ has the following symmetry properties:
\begin{equation}\label{4.10ax}
\forall {\bf f}\in L^{3/2}_{AS}(\Omega)\quad T_0{\bf f}\in
H_{AS}(\Omega),
\end{equation}
\begin{equation}\label{4.11ax}
\forall {\bf f}\in L^{3/2}_{ASwR}(\Omega)\quad T_0{\bf f}\in
H_{ASwR}(\Omega).
\end{equation}
}\end{lem} \pr The property~(\ref{4.10ax}) can be proved in the
same way as Lemma~\ref{kmpLem14.3'} and the
property~(\ref{4.11ax})  as Lemma~\ref{kmpLem14.3''}. $\qed$

\begin{lem}
\label{kmpLem14.5} {\sl The following inclusions are fulfilled:
\begin{equation}\label{4.12ax}
\forall {\bf u},{\bf v}\in H_{AS}(\Omega) \qquad ({\bf
u}\cdot\nabla){\bf v}\in L^{3/2}_{AS}(\Omega),
\end{equation}
\begin{equation}\label{4.13ax}
\forall {\bf u},{\bf v}\in H_{ASwR}(\Omega) \qquad ({\bf
u}\cdot\nabla){\bf v}\in L^{3/2}_{ASwR}(\Omega).
\end{equation}
}\end{lem}

\pr by direct calculation. $\qed$
\\

Assume that  ${\bf a}\in W^{1/2,2}(\partial\Omega)$ and let
conditions $(\ref{flux})$, (AS) (or (ASwR)\,) be fulfilled. Take
the corresponding axially symmetric functions $\bf A$, $\bf U$
from the above Lemmas. Denote ${\bf w}={\bf u}-{\mathbf U}$. Then
the problem~(\ref{NS}) is equivalent to the following one
\begin{equation}
\label{NS_1}
 \left\{\begin{array}{rcl}
-\nu \Delta{\bf w}+\big({\bf U}\cdot \nabla\big){\bf w} +
\big({\bf
w}\cdot \nabla\big){\bf w}+\big({\bf w}\cdot \nabla\big){\bf U}\\[4pt]
= - \nabla p -\big({\bf U}\cdot \nabla\big){\bf
U}\qquad \hbox{\rm in }\;\;\Omega,\\[4pt]
\div\,{\bf w}   =  0  \qquad \hbox{\rm in }\;\;\Omega,
\\[4pt]
 {\bf w}   =  0
 \qquad \hbox{\rm on }\;\;\partial\Omega.
\end{array}\right.
\end{equation}

By a {\it weak solution} of problem (\ref{NS}) we understand a
function ${\bf u}$ such that ${\bf w}={\bf u}-{\bf U}\in
H(\Omega)$ and
\begin{displaymath}
\nu\langle{\bf
w},\bfeta\rangle_{H(\Omega)}=-\int\limits_\Omega\big({\bf U}\cdot
\nabla\big){\bf U}\cdot\bfeta\,dx- \int\limits_\Omega\big({\bf
U}\cdot \nabla\big){\bf w} \cdot\bfeta\,dx
\end{displaymath}
\begin{equation}\label{4.3}
- \int\limits_\Omega\big({\bf w}\cdot \nabla\big){\bf
w}\cdot\bfeta\,dx- \int\limits_\Omega\big({\bf w}\cdot
\nabla\big){\bf U}\cdot\bfeta\,dx \qquad\forall\bfeta\in
H(\Omega).
\end{equation}

Because of Riesz representation theorem for any ${\bf w}\in
H(\Omega)$ there exists a unique function $T{\bf w}\in H(\Omega)$
such that the right-hand side of the equality~(\ref{4.3}) is
equivalent to $\langle T{\bf w},\bfeta\rangle_{H(\Omega)}$ for all
$\bfeta\in H(\Omega)$. Obviously, $T$ is a nonlinear operator from
$H(\Omega)$ to $H(\Omega)$.

\begin{lem}
\label{kmpLem14.6} {\sl The operator $T:H(\Omega)\to H(\Omega)$ is
a compact operator. Moreover, $T$ has the following symmetry
properties:
\begin{equation}\label{4.15ax}
\forall {\bf w}\in H_{AS}(\Omega)\quad T{\bf w}\in H_{AS}(\Omega),
\end{equation}
\begin{equation}\label{4.16ax}
\forall {\bf w}\in H_{ASwR}(\Omega)\quad T{\bf w}\in
H_{ASwR}(\Omega).
\end{equation}
}
\end{lem}

\pr The first statement is well known (see \cite{Lad}). The
statements about symmetry follow from the above Lemmas. $\qed$
\\

Obviously, the identity~(\ref{4.3}) is equivalent to the operator
equation in the space $H(\Omega)$:
\begin{equation}\label{4.17ax}
\nu {\bf w}=T{\bf w}.
\end{equation}

Thus, we can apply the Leray--Schauder fixed point Theorem to the
compact operators $T|_{H_{AS}(\Omega)}$ and
$T|_{H_{ASwR}(\Omega)}$. There hold the following statements.

\begin{lem} \label{kmpLem14.7}{\sl  Let the
conditions~{\rm (AS)}, $(\ref{flux})$  be fulfilled. Suppose that
all possible solutions of the equation $\nu{\bf w}=\lambda T{\bf
w}$, $\lambda\in[0,1]$, ${\bf w}\in H_{AS}(\Omega),$ are uniformly
bounded in $H_{AS}(\Omega)$. Then the problem $(\ref{NS})$ admits
at least one weak axially symmetric solution.}
\end{lem}

\begin{lem} \label{kmpLem14.9} {\sl Let the
conditions~{\rm (ASwR)}, $(\ref{flux})$  be fulfilled. Suppose
that all possible  solutions of the equation $\nu{\bf w}=\lambda
T{\bf w}$, $\lambda\in[0,1]$, ${\bf w}\in H_{ASwR}(\Omega),$ are
uniformly bounded in $H_{ASwR}(\Omega)$. Then the problem
$(\ref{NS})$ admits at least one weak axially symmetric solution
without rotation.}
\end{lem}

\subsection{On Morse-Sard and Luzin N-properties of Sobolev
functions from $W^{2,1}$} \label{SMS}

First we  recall some classical differentiability properties of
Sobolev functions.

\begin{lem}[see Proposition~1 in \cite{Dor}]
\label{kmpThDor} {\sl Let  $\psi\in W^{2,1}(\R^2)$. Then the
function~$\psi$ is continuous and there exists a set $A_{\psi}$
such that $\mathfrak{H}^1(A_{\psi})=0$, and the function $\psi$ is
differentiable (in the classical sense) at each $x\in\R^2\setminus
A_{\psi}$. Furthermore, the classical derivative at such points
$x$ coincides with $\nabla\psi(x)=\lim\limits_{r\to 0}
\dashint_{B_r(x)}{ \nabla\psi}(z)dz$, where \ $\lim\limits_{r\to
0}\dashint\nolimits_{B_r(x)}|\nabla\psi(z)-\nabla\psi(x)|^2dz=0$.}
\end{lem}

Here and henceforth we denote by $\mathfrak{H}^1$ the
one-dimensional Hausdorff measure, i.e.,
$\mathfrak{H}^1(F)=\lim\limits_{t\to 0+}\mathfrak{H}^1_t(F)$,
where $\mathfrak{H}^1_t(F)=\inf\{\sum\limits_{i=1}^\infty {\rm
diam} F_i:\, {\rm diam} F_i\leq t, F\subset
\bigcup\limits_{i=1}^\infty F_i\}$.

The next theorems have been proved recently by J. Bourgain,
M.~Korobkov and J. Kristensen \cite{korob}.

\begin{theo}
\label{kmpTh1.1}{\sl  Let  ${\mathcal D}\subset\R^2$ be a bounded
domain with Lipschitz boundary and  $\psi\in W^{2,1}({\mathcal
D})$. Then

{\rm (i)} $\mathfrak{H}^1(\{\psi(x)\,:\,x\in\bar{\mathcal
D}\setminus A_\psi\,\,\&\,\,\nabla \psi(x)=0\})=0$;

{\rm (ii)} for every $\varepsilon>0$ there exists $\delta>0$ such
that for any set $U\subset \bar{\mathcal D}$ with
$\mathfrak{H}^1_\infty(U)<\delta$ the inequality
$\mathfrak{H}^1(\psi(U))<\varepsilon$ holds.

{\rm (iii)} for $\mathfrak{H}^1$--almost all $y\in
\psi(\bar{\mathcal D})\subset \mathbb{R}$ the preimage
$\psi^{-1}(y)$ is a finite disjoint family of $C^1$--curves $S_j$,
$ j=1, 2, \ldots, N(y)$. Each $S_j$ is either a cycle in
${\mathcal D}$ $($i.e., $S_j\subset{\mathcal D}$ is homeomorphic
to the unit circle $\mathbb{S}^1)$ or it is a simple arc with
endpoints on $\partial{\mathcal D}$ $($in this case $S_j$ is
transversal to $\partial{\mathcal D}\,)$. }
\end{theo}

\begin{theo}
\label{kmpTh1.2} {\sl Let  $\mathcal D\subset\R^2$ be a bounded
domain with Lipschitz boundary and  $\psi\in W^{2,1}(\mathcal D)$.
Then for every $\varepsilon>0$ there exists an open set
$V\subset\mathbb{R}$ and a function $g\in C^1(\mathbb{R}^2)$ such
that $\mathfrak{H}^1(V)<\varepsilon$, and for each
$x\in\bar{\mathcal D}$ if $\psi(x)\notin V$ then $x\notin
A_{\psi}$, the function $\psi$ is differentiable at the point $x$,
and $\psi(x)=g(x)$, $\nabla \psi(x)=\nabla g(x)\neq 0$.}
\end{theo}

We shall say that a value  $y\in\psi(\bar{\mathcal D})$ is {\it
regular} if it satisfies the condition~(iii) of
Theorem~\ref{kmpTh1.1} and $\psi(x)\notin V$ for some $g,V$ from
Theorem~\ref{kmpTh1.2}. Note, that by above Theorems almost all
values $y\in\psi(\bar{\mathcal D})$ are regular.

 \section{Euler
equation}
\label{eueq}

\setcounter{theo}{0} \setcounter{lem}{0}
\setcounter{lemr}{0}\setcounter{equation}{0}

We will study the Euler equation under the following assumptions.

(E) {\sl Let the conditions {\rm (SO)} be fulfilled. Suppose that
some axially symmetric functions ${\bf v}\in W^{1,2}(\Omega)$ and
$p\in W^{1,3/2}(\Omega)$ satisfy the Euler system
\begin{equation} \label{2.1}\left\{\begin{array}{rcl}
\lambda_0\big({\bf v}\cdot\nabla\big){\bf v}+\nabla p & = & 0,\\[4pt]
\div{\mathbf v} & = & 0
\end{array}\right.
\end{equation}
for almost all $x\in\Omega$. Moreover, suppose that
\begin{equation}
\label{ax1}{\bf v}|_{\partial\Omega}=0.
\end{equation}}

Denote $P_+=\{(0,{x_2},{x_3}):{x_2}>0,\ {x_3}\in\R\}$, \
${\D}=\Omega\cap P_+$, $\D_j=\Omega_j\cap P_+$. Of course, on
$P_+$ the coordinates $x_2,x_3$ coincides with coordinates $r,z$.
From the conditions (SO) one can easily see that

(S${}_1$) ${\D}$ is a bounded plane domain with Lipschitz
boundary. Moreover, $C_j:=P_+\cap\Gamma_j$ is a connected set for
each $j=0,\dots,N$. In other words, the family
$\{C_j:j=0,\dots,N\}$ coincides with the family of all connected
components of the set $P_+\cap\partial{\D}$.

Then ${\bf v}$ and $p$ satisfy the following system of equations
in the plane domain~${\mathcal D}$:
\begin{equation} \label{2.1'}\left\{\begin{array}{rcl}
\frac{\partial p}{\partial z}+\lambda_0v_r\frac{\partial
v_z}{\partial r}+\lambda_0 v_z\frac{\partial
v_z}{\partial z}=0,\\[6pt]
\frac{\partial p}{\partial
r}-\lambda_0\frac{(v_\theta)^2}r+\lambda_0v_r\frac{\partial
v_r}{\partial r}+
\lambda_0v_z\frac{\partial v_r}{\partial z}=0,\\[6pt]
\frac{v_\theta v_r}r+v_r\frac{\partial v_\theta}{\partial r}+
v_z\frac{\partial v_\theta}{\partial z}=0,\\[6pt]
\frac{\partial (rv_r)}{\partial r}+\frac{\partial (rv_z)}{\partial
z}=0
\end{array}\right.
\end{equation}
(these equations are fulfilled for almost all $x\in{\mathcal
D}$\,).

The next statement  was proved in \cite[Lemma 4]{KaPi1} and in
\cite[Theorem 2.2]{Amick}.

\begin{theo}
\label{kmpTh2.3'} {\sl Let the conditions {\rm (E)} be fulfilled.
Then
\begin{equation} \label{bp2} \forall j\in\{0,\dots,N\} \ \exists\, p_j\in\R:\quad
p(x)\equiv p_j\quad\mbox{for }\Ha^2-\mbox{almost all }
x\in\Gamma_j.\end{equation} In particular, by axial symmetry,
\begin{equation}
\label{bp1}p(x)\equiv p_j\quad\mbox{for }\Ha^1-\mbox{almost all }
x\in C_j.\end{equation} }
\end{theo}

\begin{lem}[e.g., \cite{Kozono}, \cite{Neustupa}]
\label{l_est}{\sl Under conditions of Theorem~\ref{kmpTh2.3'}, the following estimate
\begin{equation} \label{bp3'}\max\limits_{i,j=0,\dots
N}|p_i-p_j|\le \delta_1\lambda_0\|\ve\|^2_{H(\Omega)}
\end{equation}
holds, where the constant $\delta_1$ depends on $\Omega$ only.}
\end{lem}

One of the main purposes of this Section is to prove the following
fact.

\begin{theo}
\label{kmpTh2.3} {\sl Under conditions of Theorem~\ref{kmpTh2.3'},
the equalities
\begin{equation}
\label{bp3}p_0=p_1=\dots=p_M\end{equation} are fulfilled.}
\end{theo}

To prove the last Theorem, we need some preparation, especially, a
version of Bernoulli Law for Sobolev case (see below
Theorem~\ref{kmpTh2.2}).

From the last equality in (\ref{2.1'}) and from~(\ref{ax1}) it
follows that there exists a stream function $\psi\in
W^{2,2}_{\loc}({\mathcal D})$ such that
\begin{equation}
\label{ax7}\frac{\partial\psi}{\partial
r}=-rv_z,\quad\frac{\partial\psi}{\partial z}=rv_r.
\end{equation}

 We have the following integral estimates: ${\bf v}\in
W^{1,2}_\loc({\mathcal D})$,
\begin{equation}
\label{ax3}\int_{{\mathcal D}}r|{\bf v}(r,z)|^2\,drdz<\infty.
\end{equation}
By identities~(\ref{ax7}), we can rewrite the last formula in the
following way:
\begin{equation}
\label{ax3'}\int_{{\mathcal
D}}\frac{|\nabla\psi(r,z)|^2}r\,drdz<\infty.
\end{equation}

Fix a point $x_*\in{\mathcal D}$. For $\varepsilon>0$ denote by
$\D_\varepsilon$ the connected component of ${\mathcal
D}\cap\{(r,z):r>\varepsilon\}$ containing~$x_*$. Since
\begin{equation}
\label{axc1}\psi\in
W^{2,2}(\D_\varepsilon)\quad\forall\varepsilon>0,
\end{equation}
by Sobolev Embedding Theorem $\psi\in C(\bar\D_{\varepsilon})$.
Hence $\psi$ is continuous at points from $\bar{\mathcal
D}\setminus O_z=\bar{\mathcal D}\setminus \{(0,z):z\in\R\}$.

Denote by $\Phi= p+\lambda_0\dfrac{|{\bf v}|^2}{2}$ the total head
pressure corresponding to the solution $({\bf v}, p)$. Obviously,
\begin{equation}
\label{axc2}\Phi\in
W^{1,3/2}(\D_\varepsilon)\quad\forall\varepsilon>0.
\end{equation}

By direct calculations one easily gets the identity
\begin{equation}
\label{2.2} v_r \frac{\partial \Phi}{\partial r}+
v_z\frac{\partial \Phi}{\partial z}=0
\end{equation}
for almost all $x\in{\mathcal D}$.

\begin{theo}
\label{kmpTh2.2} {\sl Let the conditions {\rm (E)} be valid (see
the beginning of this Section). Then there exists a set
$A_\ve\subset P_+$ such that $\Ha^1(A_\ve)=0$ and for any compact
connected\footnote{We understand the connectedness  in the sense
of general topology.} set $K\subset \bar\D\setminus O_z$,  if
\begin{equation}
\label{2.4} \psi\big|_{K}=\const,
\end{equation}
then the identities
\begin{equation}
\label{2.5'}  \Phi(x_1)=\Phi(x_2) \quad\mbox{for
 all \,}x_1,x_2\in K\setminus A_{\bf v}
\end{equation}
hold.}
\end{theo}

Theorem~\ref{kmpTh2.2} was obtained for plane case
in~\cite[Theorem~1]{korob1} (see also~\cite{kpr} for detailed
proof).

To prove Theorem~\ref{kmpTh2.2}, we need some preliminaries.

\begin{lem}
\label{pp} {\sl Let the conditions {\rm (E)} be fulfilled. Then
the inclusion
\begin{equation} \label{co1}p\in
W^{2,1}_\loc(\D)
\end{equation}
holds.}
\end{lem}

\pr Clearly, $p$ is the (unique) weak solution to the Poisson
equation
\begin{equation} \label{r2.2}\left\{\begin{array}{rcl}
\Delta p+\lambda_0\nabla{\mathbf v}\cdot\nabla{\mathbf v}^\top  & = &0\;\,\quad\hbox{\rm in }\Omega\\[4pt]
p & = & \tilde p,\quad\hbox{\rm in
}\partial\Omega,\end{array}\right.
\end{equation}
with $\tilde p=\hbox{\rm tr}\,_{\vert\partial\Omega} p\in
W^{1/3,3/2}(\partial\Omega)$. Let
$$
G(x)={\lambda_0\over 4\pi}\int_{\Omega}{(\nabla{\mathbf
v}\cdot\nabla{\mathbf v}^\top)(y)\over |x-y|} dv_y.
$$
By the results of \cite{CLMS} $\nabla{\mathbf
v}\cdot\nabla{\mathbf v}^\top$ belongs to the Hardy space ${\cal
H}^1$ so that by Calder\'on--Zygmund theorem  for Hardy's spaces
\cite{Stein} $G\in W^{2,1}(\Omega)$. Let $\bar G\in
W^{1/3,3/2}(\partial\Omega)$ be the trace of $G$ on
$\partial\Omega$ and let $p_*\in C^\infty(\Omega)$ be the solution
to the problem
\begin{equation} \label{r2.3}\left\{\begin{array}{rcl}
\Delta p_* & = &0\qquad\quad\hbox{\rm in }\Omega,\\[4pt]
p_* & = & \tilde p-\bar G \quad\hbox{\rm in }\partial\Omega.
\end{array}\right.
\end{equation}
By  the uniqueness theorem
$$
p=p_*+G(x)\in W^{2,1}_{\rm loc}(\Omega).
$$
$\qed$

\medskip

From inclusion~(\ref{co1}) it follows that ~$p_{rz}\equiv p_{zr}$
for almost all $x\in\D$. Denote $Z=\{x\in\D:v_r(x)=v_z(x)=0\}$.
Equations~(\ref{2.1'}) yield the equality
$$\frac{\partial p}{\partial z}(x)=0,\quad \frac{\partial
p}{\partial r}(x)=\lambda_0\frac{(v_\theta)^2}{r}\quad\mbox{for
almost all }x\in Z,$$ and it is easy to deduce that
\begin{equation}
\label{2.2a}\frac{\partial \Phi}{\partial z}(x)=0\quad\mbox{for
almost all }x\in\D\mbox{ such that }v_r(x)=v_z(x)=0.
\end{equation}

Consider the stream function $\psi$. From (\ref{ax1}), (\ref{ax7})
we have $\nabla \psi(x)=0$ for $\mathfrak{H}^1$-almost all
$x\in\partial{\mathcal D}\setminus O_z$. Then from the Morse-Sard
property (see Theorem~\ref{kmpTh1.1})  it follows that
$$
\mbox{for \ any\ connected\ set\ }C\subset\partial{\mathcal
D}\setminus O_z\ \exists\, \alpha=\alpha(C)\in\R :\quad
\psi(x)\equiv\alpha\ \forall x\in C.$$ Then by (S${}_1$) (see the
beginning of Section~\ref{eueq})
\begin{equation}
\label{ax8}\forall j\in\{0,\dots,N\}\ \ \exists\, \xi_j\in\R:
 \quad \psi(x)\equiv\xi_j\ \forall x\in C_j.
\end{equation}

\begin{lemr}
\label{rem_ext}{\rm Since $\nabla\psi=0$ on $\partial D\setminus
O_z$ (in the sense of traces), we can extend the function $\psi$
to the whole half-plane $P_+$:
\begin{equation}
\label{axc10} \psi(x):=\xi_0,\ \, x\in P_+\setminus
\D_0,\quad\psi(x):=\xi_j,\ \, x\in P_+\cap\bar\D_j,\ j=1,\dots,N.
\end{equation}
The functions $\ve,p,\Phi$ can be extended to $P_+$  as follows:
\begin{equation}
\label{axc10.1} \ve(x):=0,\quad x\in P_+\setminus\D,
\end{equation}
\begin{equation}
\label{axc11} p(x)=\Phi(x):=p_0,\ \, x\in P_+\setminus \D_0,\quad
p(x)=\Phi(x):=p_j,\ \, x\in P_+\cap\bar\D_j,\ j=1,\dots,N.
\end{equation}
Then the extended functions inherit the properties of the previous
ones. Namely, formulas (\ref{2.1'}), (\ref{ax7})--(\ref{2.2}),
(\ref{2.2a}) are fulfilled with $\D$, $\D_\varepsilon$ replaced by
$P_+$ and
\begin{equation}
\label{T2}P_\varepsilon:=\{(r,z):r\in[\varepsilon,\frac1\varepsilon],\
z\in[-\frac1\varepsilon,\frac1\varepsilon]\}\end{equation}
respectively.}
\end{lemr}

For $r_0>0$ denote by $L_{r_0}$ the straight line parallel to the
$z$-axis: \,$L_{r_0}=\{(r_0,z):z\in\R\}$.

Working with Sobolev functions we always assume that the "best
representatives" are chosen. The basic properties of these "best
representatives" are collected in the following

\begin{theo}
\label{kmpTh2.1} {\sl There exists a set $A_{\bf v}\subset P_+$
such that:

 {\rm (i)}\quad $ \mathfrak{H}^1(A_{\bf v})=0$;

{\rm (ii)} For  all  $x\in P_+\setminus A_{\bf v}$
\begin{displaymath}
\lim\limits_{r\to 0}\dashint\nolimits_{B_r(x)}|{\bf v}(y)-{\bf
v}(x)|^2dy=\lim\limits_{r\to
0}\dashint\nolimits_{B_r(x)}|{\Phi}(y)-{\Phi}(x)|^{3/2}dy=0,
\end{displaymath}
\begin{displaymath}
\lim\limits_{r\to
0}\frac1r\int\nolimits_{B_r(x)}|\nabla\Phi(y)|^{3/2}dy=0,
\end{displaymath}
moreover, the function $\psi$ is differentiable at $x$ and
$\nabla\psi(x)=(-rv_z(x), rv_r(x))$;

{\rm  (iii) } For all $\varepsilon >0$ there exists an open set
$U\subset \mathbb{R}^2$ such that
$\mathfrak{H}^1_\infty(U)<\varepsilon$, $A_{\bf v}\subset U$, and
the functions ${\bf v}, \Phi$ are continuous in $P_+\setminus U$;

{\rm  (iv) } For each $x_0=(r_0,z_0)\in P_+\setminus A_{\bf v}$ and
for any $\varepsilon>0$ the convergence
\begin{equation}
\label{z2}\lim\limits_{\rho\to0+}\frac1{2\rho}\Ha^1(E(x_0,\varepsilon,\rho))\to1
\end{equation}
holds, where
$$E(x_0,\varepsilon,\rho):=\{t\in(-\rho,\rho):
\int\limits_{r_0-\rho}^{r_0+\rho}\biggl|\frac{\partial
\Phi}{\partial r}(r,z_0+t)\biggr|\,dr+
\int\limits_{z_0-\rho}^{z_0+\rho}\biggl|\frac{\partial
\Phi}{\partial z}(r_0+t,z)\biggr|\,dz$$
$$+\sup\limits_{r\in[r_0-\rho,r_0+\rho]}|\Phi(r,z_0+t)-\Phi(x_0)|+
\sup\limits_{z\in[z_0-\rho,z_0+\rho]}|\Phi(r_0+t,z)-\Phi(x_0)|<\varepsilon\}.$$

{\rm  (v) } Take any function $g\in C^1(\R^2)$ and a closed set
$F\subset P_+$ such that $\nabla g\ne 0$ on $F$. Then for almost
all $y\in g(F)$ and for all the connected components $K$ of the
set $F\cap g^{-1}(y)$ the equality $K\cap A_{\ve}=\emptyset$
holds, the restriction $\Phi|_K$ is an absolutely continuous
function, and formulas~(\ref{2.1'}),\,(\ref{2.2}) are fulfilled
$\Ha^1$-almost everywhere on~$K$.}
\end{theo}

Most of these properties are from \cite{evans}. For the detailed
proof of Theorem~\ref{kmpTh2.1} see, e.g.,~\cite{kpr}. The last
property~(v) follows (by coordinate transformation, cf.
\cite[\S1.1.7]{maz'ya}) from the well-known fact that any function
$f\in W^{1,1}$ is absolutely continuous along almost all
coordinate lines. The same fact together with (\ref{2.2a}),
(\ref{2.2}) imply

\begin{lem}
\label{kmpLem2}{\sl For almost all $r_0>0$ the equality
$L_{r_0}\cap A_\ve=\emptyset$ holds, moreover, $p(r_0,\cdot)$,
$\ve(r_0,\cdot)$ are absolutely continuous functions (locally) and
\begin{equation}
\label{z1}\frac{\partial \Phi}{\partial z}(r_0,z)=0\quad\mbox{for
almost all }z\in\R\mbox{ such that }v_r(r_0,z)=0.
\end{equation}}
\end{lem}

Below we prove that for the set $A_\ve$ from Theorem~\ref{kmpTh2.1} the assertion
of Bernoulli Law (Theorem~\ref{kmpTh2.2}) holds. Before we need some lemmas.

\begin{lem}
\label{lemTh2.2} {\sl For almost all $y\in\psi( P_+)$ the equality
\begin{equation}
\label{T1}\psi^{-1}(y)\cap A_\ve=\emptyset
\end{equation}
holds, and for each continuum\footnote{By {\it continuum} we mean
a compact connected set.} $K\subset \psi^{-1}(y)$ the identities
\begin{equation}
\label{2.5}  \Phi(x_1)=\Phi(x_2) \quad\mbox{for
 all \,}x_1,x_2\in K
\end{equation}
are fulfilled.}
\end{lem}

\pr

Fix any $\varepsilon>0$ and consider a function $g\in C^1(\R^2)$
and an open set $V$ with $\mathfrak{H}^1(V)<\varepsilon$ from
Theorem~\ref{kmpTh1.2} applied to the
function~$\psi|_{P_\varepsilon}$, where the rectangle
$P_\varepsilon$ was defined by formula~(\ref{T2}). Put
$F=P_\varepsilon\setminus\psi^{-1}(V)$. Then $\psi(x)= g(x)$ and
$\nabla \psi(x)= \nabla g(x)\ne 0$ for any $x\in F$. Thus, by
Theorem~\ref{kmpTh2.1}~(v) for almost all $y\in
\psi(P_\varepsilon)\setminus V=g(F)$ and for any connected
component $K$ of the set $\{x\in P_\varepsilon: \psi(x)=y\}$ the
equality $K\cap A_\ve=\emptyset$ holds and  the restriction
$\Phi|_K$ is absolutely continuous, moreover, for any
$C^1$--smooth parametrization $\gamma:[0,1]\to K$ the
identity~(\ref{2.2}) gives
$$
[\Phi(\gamma(t))]'=\nabla
\Phi(\gamma(t))\cdot\gamma'(t)=0\quad\mbox{for }\Ha^1\mbox{-almost
all }t\in[0,1]
$$
(the last equality is valid  because  $\psi(x)=\const$ on $K$ and,
hence, $\nabla
\psi(\gamma(t))\cdot\gamma'(t)=r(-v_z(\gamma(t)),v_r(\gamma(t)))\cdot\gamma'(t)=0$).
So, we have $\Phi (x)=\const$ on $K$. In view of arbitrariness of
$\varepsilon>0$ we have proved the assertion of the Lemma. $\qed$

We need also some technical facts about continuity properties of
$\Phi$ at "good" points $x\in P_+\setminus A_\ve$.

\begin{lem}
\label{lemTh4} {\sl Let
$x_0\in P_+\setminus A_\ve$. Suppose that there exist a constant
$\sigma>0$ and  a sequence of continuums $K_j\subset P_+\setminus
A_\ve$ such that $\Phi|_{K_j}\equiv \beta_j$, $K_j\subset
B_{x_0}(\rho_j)$, $\rho_j\to 0$ as $j\to\infty$, and
$\diam(K_j)\ge\sigma \rho_j$. Then $\beta_j\to\Phi(x_0)$ as
$j\to\infty$.}
\end{lem}

\pr Without loss of generality we may assume that the projection
of each $K_j$ on the $O_r$-axis is a segment
$I_j\subset[r_0-\rho_j,r_0+\rho_j]$ of the length
$\frac12\sigma\rho_j$ (otherwise the corresponding fact is valid
for projection of $K_j$ on the $O_z$-axis, etc.) So by
Theorem~\ref{kmpTh2.1}~(iv) for any $\varepsilon>0$ we have
$I_j-\{r_0\}\cap E(x_0,\varepsilon,\rho_j)\ne\emptyset$ for
sufficiently large~$j$. Thus $|\beta_j-\Phi(x_0)|<\varepsilon$ for
sufficiently large~$j$. $\qed$

\medskip

\begin{lem}
\label{lemTh5} {\sl Suppose
for $r_0>0$ the assertion of Lemma~\ref{kmpLem2} is fulfilled,
i.e., the equality $L_{r_0}\cap A_\ve=\emptyset$ holds,
$p(r_0,\cdot)$, $\ve(r_0,\cdot)$ are absolutely continuous
functions, and formula~(\ref{z1}) is valid. Let $F\subset\R$ be a
compact set such that
\begin{equation}
\label{T1.1}\psi(r_0,z)\equiv\const\quad\mbox{for all }z\in F
\end{equation}
and
\begin{equation}
\label{T1.2}\Phi(r_0,\alpha)=\Phi(r_0,\beta) \quad\mbox{ for any
interval }(\alpha,\beta)\mbox{ adjoining }F
\end{equation}
(recall that $(\alpha,\beta)$ is called an interval adjoining~$F$
if $\alpha,\beta\in F$ and $(\alpha,\beta)\cap F=\emptyset$\,).
Then
\begin{equation}
\label{T1.3}\Phi(r_0,z)\equiv\const\quad\mbox{for all }z\in F.
\end{equation}
}
\end{lem}

\pr Take any pair $z',z''\in F$, \ $z'<z''$. On the interval
$[z',z'']$ define a function $g(z)$ by the rule
$g(z)=\Phi(r_0,z)$.  By construction, $g(\cdot)$ is an absolute
continuous functions, and from~(\ref{T1.2}) it follows that
$g(\alpha)=g(\beta)$ for any interval $(\alpha,\beta)\subset
[z',z'']$ adjoining~$F$. Since by definition the absolutely
continuous function $g(z)$ is differentiable almost everywhere and
it coincides with the Lebesgue integral of its derivative, we
obtain
\begin{displaymath}
\int_{\alpha}^{\beta} g^\prime(z)\,dz=0.
\end{displaymath}
Hence,
\begin{equation}
\label{2.7} \int_{\mu}^{\nu} g^\prime(z)\,dz=0
\end{equation}
if $\mu,\nu\in F\cap[z',z'']$ and the interval $(\mu, \nu)$
contains only a finite number of points from~$F$.

Consider now the closed set
$$\begin{array}{l}
F_\infty=\{z\in[z',z'']: \mbox{\sl in\ any\ neighborhood\ of\
the\ point}\ z \ \mbox{\sl there\ exist}\\
\\
\qquad\qquad\qquad \qquad \mbox{\sl infinitely\ many\ points\
from}\ F \}.
\end{array}
$$
It  follows from (\ref{2.7}) that
\begin{equation}
\label{2.8} \int_{[z',\,z'']\setminus F_\infty} g^\prime(z)\,dz=0.
\end{equation}
According to the~properties (ii) in Theorem~\ref{kmpTh2.1}, the
function~$\psi$ is differentiable at any point $(r_0,z)$,
$z\in(z', z'')$. From this fact and identity~(\ref{T1.1}) we
obtain $\psi_z(r_0,z)=0$ for all $z\in F_\infty$. Using
(\ref{ax7}), we can rewrite the last fact in the form
$v_r(r_0,z)=0$ for all $z\in F_\infty$. Then, in
 view of formula~(\ref{z1}), we
immediately derive
\begin{equation}
\label{2.9} \int_{ F_\infty} g^\prime(z)\,dz=0.
\end{equation}
Summing formulas (\ref{2.8}) and (\ref{2.9}), we get
\begin{displaymath}
g(z')-g(z'')=\int_{z'}^{z''} g^\prime(z)\,dz=0.
\end{displaymath}
 The last relation is equivalent to
the target equality $\Phi(r_0,z')=\Phi(r_0,z'')$. The Lemma is
proved. $\qed$
\medskip

{\bf Proof of Theorem~\ref{kmpTh2.2}}. \textsc{Step 1.} Because of
Remark~\ref{rem_ext} we can assume without loss of generality that
a continuum $K$ is a connected component of the set $\{x\in
P:\psi(x)=y_0\}$, where $y_0\in\R$, $P\subset P_+$ is a rectangle
$P:=\{(r,z):r\in[r_1,r_2],\ z\in[z_1,z_2]\}$, $r_1>0$, and
$\psi(x)\equiv \xi_0$, $\Phi(x)\equiv p_0$  for each
$x\in\partial^*P$, where we denote
$$\partial^*P=\partial P\setminus \{(r_1,z):z\in(z_1,z_2)\}.$$

Put $P^\circ=\Int P=(r_1,r_2)\times(z_1,z_2)$. For $\varepsilon>0$
denote by $K_\varepsilon$ the connected component of the compact
set $\{x\in P:\psi(x)\in[y_0-\varepsilon,y_0+\varepsilon]\}$
containing~$K$. Clearly, $K_\varepsilon\to K$ in Hausdorff metric
as $\varepsilon\to0$. By Theorem~\ref{kmpTh1.1} and
Lemma~\ref{lemTh2.2}
 for almost all $\varepsilon>0$ the set
$P^\circ\cap\partial K_\varepsilon$ is a finite disjoint union of
$C^1$-curves and functions $\psi,\Phi$ are constant on each of
these curves. From the last two sentences by topological
obviousness it follows that for each component $U_i$ of the open
set $P^\circ\setminus K$ there exists a sequence of continuums
$K^i_j\subset \bar U_i\setminus A_\ve$ such that each $K^i_j$ is a
$C^1$-curve homeomorphic to the segment $[0,1]$ or to the circle
$\mathbb S^1$, $K^i_j$ is a connected component of the set $\{x\in
P:\psi(x)=\alpha^i_j\ne y_0\}$, $\Phi|_{K^i_j}\equiv\beta^i_j$,
$K^i_j\to K\cap\partial U_i$ in Hausdorff metric as $j\to\infty$,
and for any $x\in U_i$ there exists an index $j_x$ such that $x$
and $K$ lie in the different connected components of the set
$P\setminus K^i_j$ for $j\ge j_x$. From these facts using
Lemma~\ref{lemTh4} it is easy to deduce that for any $U_i$ there
exists a limit $\beta_i=\lim_{j\to\infty}\beta^i_j$ such that
\begin{equation}
\label{T4}  \Phi(x)=\beta_i \quad\mbox{for
 all \,}x\in K\cap\partial U_i\setminus A_\ve.
\end{equation}

\textsc{Step 2.} We claim that for almost all $r_0\in(r_1,r_2)$
the identities
\begin{equation}
\label{T5}
\Phi(r_0,z')=\Phi(r_0,z'')\quad\forall\,(r_0,z'),(r_0,z'')\in K
\end{equation}
hold. Indeed, let $r_0\in(r_1,r_2)$ satisfying the assertion of
Lemma~\ref{kmpLem2} and $(r_0,z'),(r_0,z'')\in K$. Put
$F=\{z\in[z',z'']:(r_0,z)\in K\}$. From~(\ref{T4}) it follows that
$\Phi(r_0,\alpha)=\Phi(r_0,\beta)$ for any interval
$(\alpha,\beta)\subset [z',z'']$ adjoining~$F$. Thus the target
identity~(\ref{T5}) follows directly from~Lemma~\ref{lemTh5}.

\textsc{Step 3.} We claim that there exists $\beta_0\in\R$ such
that
\begin{equation}
\label{T6} \beta_i\equiv\beta_0
\end{equation}
for each component $U_i$ (see formula~(\ref{T4})). The proof of
this claim splits in two cases.

3a) \,Let $K\cap\partial^*P\ne\emptyset$. Then by construction
(see the beginning of Step~1) $y_0=\xi_0$, $K\supset\partial^*P$,
$\Phi|_{\partial^*P}\equiv p_0$, and from~(\ref{T4})--(\ref{T5})
it is easy to deduce that~$\beta_i\equiv p_0$.

3b) \,Now suppose $K\cap\partial^*P=\emptyset$. Let $U_{1}$ be a
component such that $\partial^*P\subset\partial U_{1}$. Then for
each horizontal line $L_{r_0}$ if $L_{r_0}\cap K\ne\emptyset$ then
$L_{r_0}\cap K\cap\partial U_{1}\ne\emptyset$. From the last fact
and from~(\ref{T4})--(\ref{T5}) it is easy to deduce
that~$\beta_i\equiv \beta_1$. Formula~(\ref{T6}) is proved
completely.

Now we can rewrite (\ref{T4})--(\ref{T5}) as follows:
\begin{equation}
\label{T4'}  \Phi(x)=\beta_0\quad\mbox{\, for all \,}x\in
K\cap\partial U_i\setminus A_\ve\, \mbox{ and for
 each }i,
\end{equation}
\begin{equation}
\label{T5'} \Phi(r,z)=\beta_0\quad\mbox{for
 almost all }r\in(r_1,r_2)\mbox{ and for any }(r,z)\in K,
\end{equation}
(where $\beta_0$ is equal either to $p_0$ or to $\beta_1$).

\textsc{Step 4.} We claim that
\begin{equation}
\label{T8} \Phi(x_0)=\beta_0
\end{equation}
for each $x_0\in K\setminus A_\ve$. Indeed, fix $x_0=(r_0,z_0)\in
K\setminus A_\ve$. The proof of the claim splits in two cases.

4a) \,Let there exists $\delta>0$ such that for any
$t\in(-\delta,\delta)$ the inequality
$K\cap\{(r_0+t,z):|z-z_0|\le|t|\}\ne\emptyset$ holds. Then the
equality~(\ref{T8}) follows from~(\ref{T5'}) and
the~assertion~(iv) of Theorem~\ref{kmpTh2.1}. Namely, fix
$\varepsilon>0$ and take $t\in (-\delta,\delta)\cap
E(x_0,\varepsilon,\rho)$ (this intersection is nonempty for
sufficiently small $\rho$) such that $L_{r_0+t}\cap
A_\ve=\emptyset$ and the identity~(\ref{T5'}) is fulfilled
for~$r=r_0+t$, i.e.,
\begin{equation}
\label{T5'''} \Phi(r_0+t,z)=\beta_0\quad\mbox{ for any }z\mbox{
such that }(r_0+t,z)\in K.
\end{equation}
By construction, $|t|<\rho$. By our assumption 4a) there exists a
point $(r_0+t,z_t)\in K$ such that $|z_t-z_0|\le |t|<\rho$. From
Theorem~~\ref{kmpTh2.1}~(iv) it follows that
$|\Phi(r_0+t,z_t)-\Phi(x_0)|<\varepsilon$. Using~(\ref{T5'''}), we
finally obtain $|\beta_0-\Phi(x_0)|<\varepsilon$.

4b)  \,Let the assumption 4a) be false. Then there exists a
sequence $0\ne t_k\to0$ such that
\begin{equation}
\label{TP5}
K\cap\{(r_0+t_k,z):|z-z_0|\le|t_k|\}=\emptyset.\end{equation} We
can assume without loss of generality that each segment
$\{(r_0+t_k,z):|z-z_0|\le|t_k|\}$ is contained in some $U_{i_k}$.
Denote by $Q_k$ the open squares $Q_k=(r_0-|t_k|,r_0+|t_k|)\times
(z_0-|t_k|,z_0+|t_k|)$. Then it is easy to deduce that for
sufficiently large~$k$ each set $\Cl(Q_k\cap K\cap\partial
U_{i_k})$ contains a continuum $K_k$ such that
$\diam(K_k)\ge|t_k|$. Indeed, by construction there exists
$r_k\in[r_0,r_0+t_k)$ such that $(r_k,z_0)\in\partial U_{i_k}$
(the existence of such $r_k$ follows from above inclusions
$(r_0,z_0)\in K\subset\R^2\setminus U_{i_k}$, \,$(r_0+t_k,z_0)\in
U_{i_k}$\,). Let $K_k$ be the closure of the connected component
of the set $Q_k\cap \partial U_{i_k}$ containing the point
$(r_k,z_0)$. Then $K_k\cap\partial Q_k\ne\emptyset$ (otherwise
there would be a contradiction with connectedness of~$K$). But by
assumption~(\ref{TP5}) \,$K_k$ does not intersect the segment
$\{(r_0+t_k,z):|z-z_0|\le|t_k|\}$. Hence $K_k$ intersects at least
one of other three sides of $\partial Q_k$. In each case $\diam
(K_k)\ge|t_k|$. Thus the equality~(\ref{T8}) follows
from~(\ref{T4'}) and Lemma~\ref{lemTh4}.

Now equality~(\ref{T8}) is proved for each $x_0\in K\setminus
A_\ve$. Thus the proof of Theorem~\ref{kmpTh2.2} is finished.
$\qed$.

\vspace{0.3cm}

{\bf Proof of Theorem~\ref{kmpTh2.3}.} To prove the
equalities~(\ref{bp3}), we shall use the Bernoulli law and the
fact that the axis $O_z$ is "almost" a stream line. More
precisely, $O_z$ is a singularity line for $\bf v$, $\psi$, $p$,
but it can be accurately approximated by usual stream lines (on
which $\Phi=\const$).

First of all, let us simplify
the geometrical setting.
Put
\begin{equation}
\label{axcdom} \tilde{\mathcal D}=\D\cup \bar\D_{M+1}\cup\dots\cup
\bar\D_N
\end{equation}
and consider an extension of $\psi,\Phi$ to $\tilde\D$ by formulas
of Remark~\ref{rem_ext}. Then the extended functions $\psi,\Phi$
inherit the properties of the previous ones. Namely, the Bernoulli
Law (see the assertion of Theorem~\ref{kmpTh2.2}) and
(\ref{ax3'})--(\ref{axc2})  hold with $\D$, $\D_\varepsilon$
replaced by $\tilde\D$, $\tilde\D_\varepsilon$, respectively.
Below we will use these facts only. So, we may assume, without
loss of generality, that $N=M$, i.e., that $\tilde\D=\D$ is
a~{\textit{simply connected}} plane domain.

From (\ref{ax3'}) it follows that there exists a sequence $r_i\to0+$
such that the convergence
\begin{equation}
\label{ax10} \int\limits_{L_i}|\nabla \psi|\,dz\to0\quad\mbox{as
}i\to+\infty
\end{equation}
holds for  lines $L_i=\{(r,z)\in\bar{\mathcal
D}:r=r_i\}$. Fix a point $x_0\in \D$ and denote by ${\mathcal D}^i$
the connected component of the open set $\{(r,z)\in{\mathcal
D}:r>r_i\}$ containing~$x_0$. Obviously, for sufficiently large $i$
the open set ${\mathcal D}^i$ is a~simply connected plane
domain with a Lipschitz boundary, $\psi\in W^{2,1}({\mathcal
D}^i)\subset C(\bar{\mathcal D}^i)$. We
also have
\begin{equation}
\label{axc13}\partial{\mathcal D}^i\setminus
L_i=C_0^i\cup\dots\cup C_M^i,\end{equation}
\begin{equation}
\label{axc14}C_j^i\cap L_i\ne\emptyset,\quad
j=0,\dots,M,\end{equation}
where $C_j^i=C_j\cap\{(r,z)\in{\mathcal
D}:r\ge r_i\}$, $j=0,\dots,M$. Then from~(\ref{ax8}) and
(\ref{ax10}) we conclude that
\begin{equation}
\label{ax11}
\diam(\psi(\partial\D^i))=\sup\limits_{x,y\in\partial\D^i}|\psi(x)-\psi(y)|\to0.
\end{equation}
In particular, $\xi_0=\dots=\xi_M$, i.e.,
\begin{equation}
\label{ax11'} \psi|_{P_+\cap\partial{\mathcal
D}}\equiv\xi_0\equiv\psi|_{\partial{\mathcal D}^i\setminus
L_i},\quad \sup\limits_{x\in\partial\D^i}|\psi(x)-\xi_0|\to0.
\end{equation}

Our plan for the rest part of the proof is as follows. First, we
prove that for any $x\in P_+\cap\bar{\mathcal D}$ there exists a
set $U(x)$ such that
\begin{equation*}
x\in U(x)\subset  P_+\cap\bar{\mathcal D},\quad O_z\cap\partial
U(x)\ne\emptyset,
\end{equation*}
\begin{equation}
\label{ax14} \psi|_{P_+\cap\partial U(x)} \equiv\xi_0,
\end{equation}
\begin{equation}
\label{ax16}  \exists \,\beta(x)\in\R:\quad \Phi(y)=\beta(x)\ \
\forall y\in P_+\cap(\partial U(x))\setminus A_{\bf v}.
\end{equation}
Notice that $\psi|_{P_+\cap\partial U(x)}=\xi_0$ does not depend
on $x$, while $\Phi|_{P_+\cap\partial U(x)}=\beta(x)$ can a~priory
depend on $x$. However, finally we prove that $\beta(x)\equiv p_0$
for all $x\in P_+\cap \bar{\mathcal D}$. This fact will easily
imply the target equalities~(\ref{bp3}).

On $\bar{\mathcal D}^i$ define an equivalence relation by the rule
$x\sim_iy\Leftrightarrow\exists$ a
conti\-nuum\footnote{By {\it continuum} we mean a
compact connected set.} $K\subset\bar{\mathcal D}^i$ such that
$\psi|_{K}\equiv\const$ and both $x,y$ do not belong to the
unbounded connected component of the open set $\R^2\setminus K$. By
$U_i(x)$ denote the corresponding class of equivalence. Illustrate
this definition by some examples.

($\mbox{I}_\sim$) \ If  $K\subset\bar{\mathcal D}^i$ is a continuum
and $\psi|_K=\const$, then $x\sim_iy \ \ \forall
x,y\in K$.

($\mbox{II}_\sim$) If $K\subset\bar{\mathcal D}^i$ is homeomorphic
to the circle and $\psi|_K\equiv\const$, then $x\sim_iy \ \
\forall x,y\in U$, where $U$ is a bounded domain such that
$\partial U=K$.

For each $x\in\bar{\mathcal D}^i$ the following properties of the relation~$\sim_i$  hold
(for the proof of them, see Appendix).

($\mbox{III}_\sim$) \  $U_i(x)\subset U_{i+1}(x)$ and each $U_i(x)$ is a compact set.

($\mbox{IV}_\sim$)  \ The set $U_i(x)$ is connected.

($\mbox{V}_\sim$) \ $\psi|_{\partial U_i(x)}\equiv\const$.

($\mbox{VI}_\sim$) \ The set $\R^2\setminus U_i(x)$ is connected.

($\mbox{VII}_\sim$) \ The set $\partial U_i(x)$ is connected.

($\mbox{VIII}_\sim$) \ The formula
\begin{equation}
\label{ax13} L_i\cap \partial U_i(x)\ne\emptyset
\end{equation}
holds.

For $x\in\bar{\mathcal D}\setminus O_z$ put $U(x)=\bigcup\limits_i
U_i(x)$. Because of topological obviousness
\begin{equation}
\label{ax15-'} \forall y\in P_+\cap\partial U(x)\ \exists\ a\
sequence\ \partial U_i(x)\ni y_i\to y.
\end{equation}
Then from ($\mbox{V}_\sim$), (\ref{ax11})--(\ref{ax11'}) and
(\ref{ax13}) we conclude that the identity~(\ref{ax14}) holds.

From the Bernoulli Law (see
Theorem~\ref{kmpTh2.2}) it follows that
\begin{equation}
\label{ax15} \forall x\in P_+\cap \bar{\mathcal D}\  \ \exists\,
\beta_i(x):\ \ \Phi(y)=\beta_i(x) \ for\  all\ y\in
\partial U_i(x)\setminus A_{\bf v}.
\end{equation}
Fix any point $y_*\in P_+\cap \partial U(x)\setminus A_{\bf v}$
and $j$ such that $y_*\in\bar{\mathcal D}_{j}\setminus L_{j}$. By
construction (see the properties
($\mbox{VII}_\sim$)--($\mbox{VIII}_\sim$), (\ref{ax15-'})\,) there
exists a sequences of continuums $K_i\subset \bar{\mathcal
D}_{j}\cap\partial U_i(x)$ and points  $ y_i\in K_i$ such that
$K_i\cap L_j\ne\emptyset$ for all sufficiently large $i$, $y_i\to
y_*$, and $K_i$ converges to some set~$K$ with respect to the
Hausdorff metric as $i\to\infty$. Hence $y_*\in K$, $K$ is a
compact connected set, $\psi|_K\equiv \xi_0=\const$, and $K\cap
L_j\ne\emptyset$. Consequently,
\begin{equation}
\label{ax15'''}\diam K>0.
\end{equation} Again by the Bernoulli Law  it follows that
\begin{equation}
\label{ax15''} \exists\, \beta\in\R:\  \quad\Phi(y)=\beta \ for\
all\ y\in K\setminus A_{\bf v}.
\end{equation}
Then from (\ref{ax15})--(\ref{ax15'''}), the connectedness of $K,
K_i$,  and from the continuity properties of $\Phi$ (see
Theorem~\ref{kmpTh2.1}\,(iii)~) we conclude that the convergence
$$
\lim\limits_{i\to\infty}\beta_i(x)=\beta
$$
holds. In particular,
$$
\Phi(y_*)=\lim\limits_{i\to\infty}\beta_i(x).
$$
Because the right-hand side of the last equality does not depend
on the choice of $y_*\in P_+\cap\partial U(x)\setminus A_{\bf
v}$, we have proved the identities (\ref{ax16}) with
$\beta(x)=\lim\limits_{i\to\infty}\beta_i(x)$.

Now let $r_0>0$ satisfying the assertion of Lemma~\ref{kmpLem2}
and the~conditions $(r_0,z'),(r_0,z'')\in P_+\cap\partial
{\mathcal D}$, \ $\{(r_0,z):z\in(z',z'')\}\subset\D$. To finish
the proof of the theorem, we need to show that
\begin{equation}
\label{2.6} \Phi(r_0,z')=\Phi(r_0,z'').
\end{equation}
Put $$F=\{z\in[z',z'']:\ (r_0,z)\in
\partial U((r_0,z))\}.
$$
Then by construction $z',z''\in F$ and the set $F$ is compact.
Indeed, denote $x'=(r_0,z')$, $x''=(r_0,z'')$. Since $U(x')\subset
P_+\cap \bar{\mathcal D}$, we have $\partial\D\ni x'\notin\Int
U(x')$, consequently, $x'\in\partial U(x')$. Analogously, $x''\in
\partial U(x'')$, i.e., $z',z''\in F$. Further, let $F\ni z_k\to
z_0$. Denote $x_k=(r_0,z_k)$. Then $x_k\in\partial U(x_k)$,
$x_k\to x_0=(r_0,z_0)$. Of course, $x_0\notin \Int U(x_0)$
(otherwise $x_k\in \Int U(x_0)=\Int U(x_k)$ for large $k$\,).
Therefore $x_0\in\partial U(x_0)$, i.e., $z_0\in F$. So, we prove
that $z',z''\in F$ and that the set $F$ is compact.

Now from (\ref{ax14})--(\ref{ax16}) the
identities~(\ref{T1.1})--(\ref{T1.2}) hold. Thus by
Lemma~\ref{lemTh5} we have the target equality~(\ref{2.6}). $\qed$

\medskip

In particular, during the last proof we established the following
assertion.

\begin{lem} \label{kmp_b}{\sl
Assume that the conditions {\rm (E)} be fulfilled. Let $K_i$ be a
sequence of compact sets with the following properties:
$K_i\subset\bar{\mathcal D}\cap P_+$, $\psi|_{K_i}=\const$, and
let there exist $x_i,y_i\in K_i$ such that $\dist(x_i,O_z)\to0$,
$\dist(y_i,O_z)\nrightarrow0$. Then there exist $\beta_i\in\R$
such that $\Phi(x)\equiv \beta_i$ $\forall x\in K_i\setminus
A_{\bf v}$ and $\beta_i\to p_0$ as $i\to\infty$. }
\end{lem}

Let $U\subset\R^2$ be a domain with Lipschitz boundary. We say
that the function $f\in W^{1,s}({U})$ satisfies a {\it weak
one-side maximum principle locally} in ${U}$, if
\begin{equation}
\label{2.13} \mathop{\hbox{\rm ess}\,\hbox{\rm
sup}}_{x\in{U}^\prime}\,f(x)\leq \mathop{\hbox{\rm ess}\,\hbox{\rm
sup}}_{x\in\partial{U}^\prime}\,f(x)
\end{equation}
holds for any strictly interior subdomain ${U}^\prime$
($\bar{U}\,^\prime\subset{U})$  with the boundary
$\partial{U}^\prime$ not containing singleton connected
components. (In (\ref{2.13}) negligible sets are the sets of
2--dimensional Lebesgue measure zero in the left {\sl esssup}, and
the sets of 1--dimensional Hausdorff measure zero in the right
{\sl esssup}.)

If (\ref{2.13}) holds for any ${U}^\prime \subset{U}$ (not
necessary strictly interior) with the boundary
$\partial{U}^\prime$ not containing singleton connected
components, then we say that $f\in W^{1,s}({U})$
 satisfies a {\it  weak one-side maximum principle globally} in ${U}$
(in particular,  we can take ${U}^\prime={U}$ in (\ref{2.13})).

\begin{theo}
\label{kmpTh2.4} {\sl Let the conditions {\rm (E)} be fulfilled.
Assume that there exists a sequence of functions $\{\Phi_\mu\}$
such that $\Phi_\mu\in W^{1,s}_{\loc}({\mathcal D})$ and
$\Phi_\mu\rightharpoonup\Phi$ weakly in $W^{1,s}_{\loc}({\mathcal
D})$ for some $s\in (1,2)$. If all $\Phi_\mu$ satisfy the weak
one--side maximum principle locally in ${\mathcal D}$, then
\begin{equation}
\label{bp4}\esssup\limits_{x\in{\mathcal
D}}\Phi(x)\le\max\limits_{j=0,\dots,N}p_j.
\end{equation}}
\end{theo}

\pr Let the conditions of Theorem~\ref{kmpTh2.4} be fulfilled.
Then from \cite[Theorem~2]{korob1} (see also \cite{kpr} for more
detailed proof) it follows that
$$
\begin{array}{l}
(*)\ for\ any\ subdomain\ U\subset \D\ such\ that\ \bar U\cap
O_z=\emptyset\ the\ function\\
\quad\Phi|_{\bar U}\  satisfies\ the\ weak\ one-side\ maximum\
principle\ globally.
\end{array}
$$

To prove the estimate~(\ref{bp4}) in the whole domain~$\D$ we will
use the same methods as in the proof of Theorem~\ref{kmpTh2.3}.
First of all, we simplify the situation: as above define the
domain~$\tilde\D$ by equality~(\ref{axcdom}) and extend the
functions $\psi,\Phi$ to $\tilde\D$ by
formulas~(\ref{axc10})--(\ref{axc11}). The extended functions
$\psi,\Phi$ inherit the properties of the previous ones. Namely,
(\ref{ax3'})--(\ref{axc2}) and the Bernoulli Law (see
Theorem~\ref{kmpTh2.2}) hold with $\D$, $\D_\varepsilon$ replaced
by $\tilde\D$, $\tilde\D_\varepsilon$, respectively. Moreover, the
 maximum property~(*) holds with $\D$ replaced
by $\tilde\D$. Since in the proof below we will use only these
facts, we may assume without loss of generality that $N=M$, i.e.,
$\tilde\D=\D$ is {\sl a~simply connected} plane domain.

Suppose the assertion of the Theorem is false. Then there exists
a~point $x_*\in\mathcal D\setminus A_{\bf v}$ such that
\begin{equation}
\label{bp5}\Phi(x_*)=p_*>\max\limits_{j=0,\dots,N}p_j.
\end{equation}

Take a sequences  of numbers $r_i\to +0$,  the corresponding lines
$L_i$ and domains $\mathcal D^i$ from the proof of
Theorem~\ref{kmpTh2.3} (in particular, the formula~(\ref{ax10})
holds). Denote by $K^*_i$ the connected component of the level set
$\{x\in\bar{\mathcal D}^i\,:\,\psi(x)=\psi(x_*)\}$
containing~$x_*$. By Bernoulli Law
\begin{equation}
\label{bp6}\Phi(x)=p_*\quad{ for\ all\ i\ and\ for\ all\ }x\in
K^*_i\setminus A_\ve.
\end{equation}

We have two possibilities:

(I) $K^*_i\cap L_i\ne\emptyset$ for all $i$. Then, by
Lemma~\ref{kmp_b},  \,$p_*=p_0$, and we obtain a contradiction
with the assumption~(\ref{bp5}).

(II) There exists $i_0$ such that $K^*_{i_0}\cap
L_{i_0}=\emptyset$. Then the sequence $K^*_i$ stabilizes after $i=
i_0$, i.e.,
\begin{equation}
\label{bp7}K^*_i=K^*_{i_0},\quad K^*_{i}\cap L_i
=\emptyset\quad\mbox{ for all }i\ge i_0.
\end{equation}
Denote $K^*=K^*_{i_0}$. Then by  construction
\begin{equation}
\label{bp8}K^*\cap \partial\mathcal D^i=\emptyset\quad\mbox{ for all }i\ge i_0.
\end{equation}
Now consider the family of sets $U_i(x_*)$ defined during the
proof of Theorem~~\ref{kmpTh2.3}. From~(\ref{ax15}) it follows
that
\begin{equation}
\label{bpax16}  \Phi(y)=\beta_i(x_*)\quad\forall y\in
\partial U_i(x_*)\setminus A_{\bf v},
\end{equation}
where
\begin{equation}
\label{bpax18} \lim\limits_{i\to\infty}\beta_i(x_*)=p_0
\end{equation}
(the last convergence follows from Lemma~\ref{kmp_b}).
Take $i_1\ge i_0$ sufficiently large such that
\begin{equation}
\label{bpax20} \beta_i(x_*)<p_*\quad\mbox{ for all }i\ge i_1.
\end{equation}
Denote $U=\Int U_{i_1}(x_*)$. By construction,
\begin{equation}
\label{2.13bp} \mathop{\hbox{\rm ess}\,\hbox{\rm
sup}}_{x\in{U}}\,\Phi(x)\ge p_*>
\beta_{i_1}(x_*)=\mathop{\hbox{\rm ess}\,\hbox{\rm
sup}}_{x\in\partial{U}}\,\Phi(x).
\end{equation}
But the last inequalities contradict the assertion~(*). The proof is
complete.

$\qed$

\section{The proof of Existence Theorem}
\label{poet}
\setcounter{theo}{0} \setcounter{lem}{0}
\setcounter{lemr}{0}\setcounter{equation}{0}

Consider first the axially symmetric case with possible rotation.
According to Lemma \ref{kmpLem14.7},  in order to prove the
existence of the solution to problem~(\ref{NS}) it is enough to
show that all possible solutions to the operator equation
\begin{equation}
\label{opereq}
\nu\w=\lambda T\w, \ \ \lambda\in[0,1], \ \w\in H_{AS}(\Omega)
\end{equation}
are uniformly bounded in $ H_{AS}(\Omega)$.  We shall prove this
estimate by contradiction,  following the well-known argument of
J.~Leray \cite{Leray} (this argument was used also by many other
authors, e.g. \cite{Lad1}, \cite{Lad}, \cite{KaPi1}, \cite{Amick},
 see also \cite{kpr}).

Suppose that the solutions to~(\ref{opereq}) are not uniformly
bounded in $ H_{AS}(\Omega)$. Then there exists a sequence of
functions~$\w_k\in H_{AS}(\Omega)$ such that $\nu{\bf
w_k}=\lambda_k T{\bf w_k}$ with $\lambda_k\in[0,1]$ and
$J_k=\|\w_k\|_{H(\Omega)}\to\infty$. Note that $\w_k$ and the
corresponding axially symmetric pressures $p_k\in
L^2_{AS}(\Omega)$ satisfy the following integral identity
\begin{displaymath}
\nu\int\limits_\Omega\nabla{\bf
w}_k\cdot\nabla\bfeta\,dx=-\lambda_k\int\limits_\Omega\big({\bf U}\cdot
\nabla\big){\bf U}\cdot\bfeta\,dx- \lambda_k\int\limits_\Omega\big({\bf
U}\cdot \nabla\big){\bf w}_k \cdot\bfeta\,dx
\end{displaymath}
\begin{equation}\label{identity}
- \lambda_k\int\limits_\Omega\big({\bf w}_k\cdot \nabla\big){\bf
w}_k\cdot\bfeta\,dx- \lambda_k\int\limits_\Omega\big({\bf w}_k\cdot
\nabla\big){\bf U}\cdot\bfeta\,dx+\int\limits_\Omega p_k{\rm div}\,\bfeta\,dx
\end{equation}
for any $\bfeta\in \wotwo(\Omega)$. Here $\U$ is an axially
symmetric solution to the Stokes problem~ (see
Lemmas~\ref{kmpLem14.3}--\ref{kmpLem14.3'}).

Denote $\u_k=\w_k+\U$, $\hu_k=\frac{1}{J_k}\u_k$,
$\hw_k=\frac1{J_k}\w_k$, $\hp_k=\frac{1}{J^2_k}p_k$. Then
$\|\hw_k\|_{H(\Omega)}=1$ and the following estimates
\begin{displaymath}
\|\widehat p_k\|_{L^{2}(\Omega)}\leq \const, \quad \|\widehat
p_k\|_{W^{1,3/2}(\Omega^\prime)}\leq \const
\end{displaymath}
hold for any $\bar\Omega\,^\prime\subset\Omega$ (the detailed
proof of the above estimates see, for example, in~\cite{kpr}).
Extracting a subsequences, we can assume without loss of
generality that
\begin{equation}
\label{exc1} \lambda_{k}\to\lambda_0\in[0,1],
\end{equation}
\begin{equation}
\label{exc3} \hu_{k}\rightharpoonup \ve\in H(\Omega)\quad{ weakly\
in\ }W^1_2(\Omega),
\end{equation}
\begin{equation}
\label{exc3'} \hp_{k}\rightharpoonup p\in
W_{loc}^{1,3/2}(\Omega)\cap L^2(\Omega)\quad{ weakly\ in\
}L^2(\Omega)\ { and\ in\ }W^{1,3/2}_{loc}(\Omega).
\end{equation}
\\
Multiplying the integral identity (\ref{identity}) an arbitrary
fixed $\bfeta\in\wotwo(\Omega)$ by $J_{k}^{-2}$ and passing to a
limit as $k\to\infty$, yields that the limit functions ${\bf v}$
and $p$ satisfy the Euler equations
\begin{equation}
\label{Euler}\left\{\begin{array}{rcl}
\lambda_0\big({\bf v}\cdot\nabla\big){\bf v}+\nabla p & = & 0,\\[4pt]
\div{\mathbf v} & = & 0
\\[4pt]
{\mathbf v}|_{\partial\Omega} & = & 0
\end{array}\right.
\end{equation}
(the details of the proof see, for example, in~\cite{kpr}). From
equations (\ref{Euler}) and from inclusions
(\ref{exc3}),~(\ref{exc3'}) it follows that $p\in
W^{1,3/2}(\Omega)$. Thus the assumptions~(E) from the beginning of
the Section~\ref{eueq} are fulfilled. Moreover,
$\|\ve\|_{H(\Omega)}\le1$.

Now, taking in (\ref{identity}) $\bfeta=J_{k}^{-2}\w_{k}$ we get
\begin{equation}\label{4.7}
\nu\int\limits_\Omega|\nabla\widehat{\bf
w}_{k}|^2\,dx=\lambda_{k}\int\limits_\Omega\big(\widehat{\bf
w}_{k}\cdot\nabla\big)\widehat{\bf w}_{k}\cdot{\bf U}\,dx+
J_{k}^{-1}\lambda_{k}\int\limits_\Omega\big({\bf
U}\cdot\nabla\big)\widehat{\bf w}_{k}\cdot{\bf U}\,dx
\end{equation}
Using the compact embedding $H(\Omega)\hookrightarrow
L^r(\Omega),$ $r<6$, we can pass to a limit as $k\to\infty$ in equality
(\ref{4.7}). As a result we obtain
\begin{equation}\label{4.8}
\nu=\lambda_0\int\limits_\Omega\big(\widehat{\bf
v}\cdot\nabla\big)\widehat{\bf v}\cdot{\bf U}\,dx.
\end{equation}
From the last formula and Euler equation~(\ref{Euler}), we derive
\begin{equation}
\label{4.8'} \nu=-\int\limits_\Omega\nabla p\cdot{\bf U}\,dx=
-\int\limits_\Omega\div (p\,{\bf
U})\,dx=-\int\limits_{\partial\Omega} p \,{\bf a}\cdot{\bf n}\,dS.
\end{equation}
Because of (\ref{bp2}) the last equality could be rewritten in the
following equivalent form
\begin{equation}
\label{exc5}\sum\limits_{j=0}^N p_j\F_j=-\nu.
\end{equation}
Now using (\ref{flux}) and (\ref{bp3}) from (\ref{exc5}) we derive
\begin{equation}
\label{exc6}p_0\sum\limits_{j=0}^M\F_j+\sum\limits_{j=M+1}^N
p_j\F_j= \sum\limits_{j=M+1}^N \F_j(p_j-p_0)=-\nu.
\end{equation}

Consider, first, the case (\ref{flux^2}).  If the condition
(\ref{flux^2}) is fulfilled with
$\delta=\frac1{\delta_1(N-M)}\nu$, where $\delta_1$ is a constant
from~Lemma~\ref{l_est}, then from~(\ref{exc6}) and (\ref{bp3'}) it
follows a contradiction (recall that $\|{\bf v}\|_{H(\Omega)}\le
1,\, \lambda_0\in[0,\,1]$). Thus, the proof the case
(\ref{flux^2}) is complete.

Consider now the case when condition (\ref{flux^1}) is fulfilled.
Then the equality~(\ref{exc6}) takes the following form:
\begin{equation}
\label{exc7}
\F_N(p_0-p_N)=\nu.\end{equation}
From (\ref{flux^1}), (\ref{exc7}) it follows that
\begin{equation}
\label{exc8}
p_0>p_N.
\end{equation}
Consider the identity
$$
\div\big(x p+\lambda_0({\bf v}\cdot x){\bf
v}\big)=\big(x\cdot\nabla  p+x\cdot\lambda_0({\bf
v}\cdot\nabla){\ve}\big)+3
p+\lambda_0|{\ve}|^2=
$$
$$
=3\big( p+\frac{\lambda_0}{2}|{\bf
v}|^2\big)-\frac{\lambda_0}{2}|{\ve}|^2=3
\Phi-\frac{\lambda_0}{2}|{\ve}|^2.
$$
 Integrating the above identity by parts in $\Omega$, we get
$$
3\int\limits_\Omega \Phi
dx-\frac{\lambda_0}{2}\int\limits_\Omega|{\bf
v}|^2dx=\int\limits_{\partial\Omega} p \, (x\cdot{\bf
n})dS=
$$
$$
= p_0 \int\limits_{\Gamma_0}(x\cdot{\bf n})dS+
p_0\sum\limits_{j=1}^{N-1} \int\limits_{\Gamma_j}(x\cdot{\bf
n})dS+ p_N\int\limits_{\Gamma_N}(x\cdot{\bf n})dS
$$
$$
= p_0 \int\limits_{\Omega_0}\div\, xdx- p_0\sum\limits_{j=1}^{N-1}
\int\limits_{\Omega_j}\div\, x dx -
p_N\int\limits_{\Omega_N}\div\, x dx=
$$
$$
=3 p_0\big(|\Omega_0|-\sum\limits_{j=1}^{N-1} |\Omega_j|\big)-3
p_N|\Omega_N|=3 p_0|\Omega| +3( p_0- p_N)|\Omega_N|.
$$
Hence,
\begin{equation}
\label{exc22}
\int\limits_\Omega \Phi dx\geq\int\limits_\Omega
\Phi dx-\frac{\lambda_0}{6}\int\limits_\Omega|{\bf
v}|^2dx= p_0|\Omega| +( p_0-
p_N)|\Omega_N|.
\end{equation}
The total head pressures $\Phi_k=p_k+\frac{\lambda_k}2|{\bf
u}_k|^2$ for the Navier--Stokes system~(\ref{NS}) satisfy the
equations
$$
\nu \Delta\Phi_k-\lambda_k{\bf u}_k\cdot\nabla\Phi_k=\nu
|\,\curl\,\bf u_k\,|^2\geq 0.
$$
Hence it is well known (see, e.g., \cite{Mi2}) that $\Phi_{k}$
satisfy the one-side maximum principle locally in $\Omega$. Denote
$\widehat\Phi_{k}=\frac{1}{J_{k}^2}\Phi_{k}$. From
(\ref{exc3})--(\ref{exc3'}) and from the symmetry assumptions it
follows that the sequence $\{\widehat\Phi_{k}\}$ weakly converges
to $ \Phi=p+\dfrac{\lambda_0}{2}|\ve|^2$ in the space
$W^{1,3/2}_{loc}({\mathcal D})$. Therefore, by Theorem
\ref{kmpTh2.4},
\begin{equation}
\label{bpex4}\esssup\limits_{x\in\Omega}\Phi(x)=\esssup\limits_{x\in{\mathcal
D}}\Phi(x)\le\max\limits_{j=0,\dots,N}p_j=p_0
\end{equation}
(the last equality follows from the conditions~$N=M+1$ and (\ref{exc8})\,).
Then it follows from (\ref{exc22}) that
$$
 p_0|\Omega| +( p_0- p_N)|\Omega_N|\leq
 p_0|\Omega|\quad \Leftrightarrow\quad
p_0\leq p_N,
$$
and we obtain the contradiction with (\ref{exc8}), which proves
Theorem in the case of condition (\ref{flux^1}).

If the boundary value~$\mathbf{a}$ is axially symmetric without
rotation, the proof of Theorem~\ref{kmpTh4.2} is just the same as
in the first part; we need only to use Lemma~\ref{kmpLem14.9}
instead of Lemma~\ref{kmpLem14.7}. $\qed$

\section{Appendix}

Let us prove the topological properties
($\mbox{III}_\sim$)--($\mbox{VIII}_\sim$) of the equivalence
class~$U_i(x)$, $x\in\bar{\mathcal D}^i$, which were used in the
proof of Theorem~\ref{kmpTh2.3}.

($\mbox{III}_\sim$) \  Indeed, if $U_i(x)\ni y_j\to y$, then by
definition there exists a sequence of continuums $K_j$ such that
$\psi|_{K_j}=\const$ and $x,y_j$ do not belong to the unbounded
connected component of the set~$\R^2\setminus K_j$. Without loss
of generality we may assume that  $K_j$ converge with respect to
the Hausdorff metric to the set~$K$. Then $K$ is a continuum,
$\psi|_K=\const$, and it is easy to see that neither $x$ nor $y$
belongs to the unbounded connected component of the open
set~$\R^2\setminus K$.

($\mbox{IV}_\sim$)  \ Fix any $y\in U_i(x)$. Take the
corresponding set~$K$ from the definition of~$x\sim_i y$. Then
$K\subset\bar{\mathcal D}^i$ is a compact connected set such that
$\psi|_{K}\equiv\const$ and both $x,y$ do not belong to the
unbounded connected component of the open set $\R^2\setminus K$.
Denote by $V_j$ the family of connected components of the open set
$\R^2\setminus K$. Let $V_0$ be an unbounded component. Since the
domain ${\mathcal D}^i$ is simply connected, we have $\bar
V_j\subset\bar{\mathcal D}^i$ for each $j\ne0$. Hence by
definition of $\sim_i$ we obtain $\bar V_j\subset U_i(x)$ for each
$j\ne0$. By construction, each set $K$, $\bar V_j$ is connected
and $K\cap \bar V_j\ne \emptyset$. From these facts we conclude
that the set $S_y=K\cup\biggl(\bigcup\limits_{j\ne0}\bar
V_j\biggr)$ is connected and the inclusions $\{x,y\}\subset
S_y\subset U_i(x)$ hold. The last assertion and arbitrariness of
$y\in U_i(x)$ imply the connectedness of~$U_i(x)$.

($\mbox{V}_\sim$) \ To prove the property $\psi|_{\partial
U_i(x)}=\const$, we may assume, without loss of generality, that
$x\in \partial U_i(x)$. Fix any $y\in \partial U_i(x)$. Take the
corresponding set~$K$ from the definition of~$x\sim_i y$ and the
sets $V_j$ from the proof of property~($\mbox{IV}_\sim$). Then it
is easy to see that
\begin{equation}
\label{appp1}x,y\in K.
\end{equation}
Indeed, if for example
$y\notin K$, then $y\in V_j$ for some $j\ne0$.  But by
construction $V_j$ is an open set and $V_j\subset U_i(x)$. These
facts contradict the assumption $y\in
\partial U_i(x)$. This proves the inclusion~(\ref{appp1}).
From~(\ref{appp1}) and  the assumption~$\psi|_{K}\equiv\const$ we
obtain the required equality~$\psi(y)=\psi(x)$.

Using similar elementary arguments, it is easy to prove the next
two  properties ($\mbox{VI}_\sim$)--($\mbox{VII}_\sim$).
Therefore, we shall prove in detail only the  last
property~($\mbox{VIII}_\sim$).

($\mbox{VIII}_\sim$) \ Suppose the formula~(\ref{ax13}) is not
true, i.e.,
\begin{equation}
\label{ax19} L_i\cap \partial U_i(x)=\emptyset.
\end{equation}
Hence,
\begin{equation}
\label{ax19'} L_i\cap U_i(x)=\emptyset.
\end{equation}
Let $\psi(y)\equiv c_0$ for all $y\in \partial U_i(x)$. Fix
$y_0\in \partial U_i(x)$. From properties ($\mbox{I}_\sim$),
($\mbox{V}_\sim$), ($\mbox{VII}_\sim$) it follows that
\begin{equation}
\label{ax19''} \partial U_i(x)\subset K_0 \subset U_i(x),
\end{equation}
where we denote by  $K_0$ the connected component of the level set
$\{y\in\bar{\mathcal D}^i:\psi(y)=c_0\}$ containing the
point~$y_0$.

By construction, the closure of each connected component $\tilde
C$ of the set $(\partial\D^i)\setminus L_i$ intersects the line
$L_i$ and $\psi|_{\tilde C}\equiv\const$ \ (see the~formulas
(\ref{axc13})--(\ref{axc14}), (\ref{ax11'})\,). Hence, the
conditions~(\ref{ax19'})--(\ref{ax19''}) imply the assertion
\begin{equation} \label{ax19'''} K_0\cap\partial\D^i=
U_i(x)\cap\partial\D^i=\emptyset.
\end{equation}
Take a sequence $0<\delta_j\to 0$ such that each value
$c_0+\delta_j, c_0-\delta_j$ is regular from the viewpoint of
Morse-Sard Theorem (see Theorem~\ref{kmpTh1.1}~(iii)). Denote by
$B_j$ the connected component of the level
set~$\{y\in\bar{\mathcal
D}^i:\psi(y)\in[c_0-\delta_j,c_0+\delta_j]\}$ containing~$K_0$.
Then for sufficiently large $j$ the boundary $\partial B_j$
consists of finite disjoint family of $C^1$--cycles in ${\mathcal
D}^i$ (it follows from the formula~(\ref{ax19'''}) and from the
evident convergence $\sup\limits_{y\in B_j}\dist(y,K_0)\to0$\,).

Denote by $K_j\subset\partial B_j$ the cycle separating the
set~$B_j$ from infinity, and denote by $U_j$ the bounded domain
such that $\partial U_j=K_j$. Then by construction
$\psi|_{K_j}\equiv\const$, $K_0\cap K_j=\emptyset$, and
$K_0\subset U_j$. Consequently,
\begin{equation}
\label{ax20} U_i(x)\subsetneqq U_j.
\end{equation}
On the other hand, by property~($\mbox{II}_\sim$) \ all points of
$U_j$ are $\sim_i$ equivalent. The last assertion contradicts the
formula~(\ref{ax20}) and the definition of~$U_i(x)$. The
property~(\ref{ax13}) is proved.

\medskip

\section*{Acknowledgements}
$\quad\,\,$The authors are deeply indebted to V.V.~Pukhnachev for valuable
discussions.

The research of M. Korobkov was supported by the Russian
Foundation for Basic Research (project No.~11-01-00819-a) and by
the Research Council of Lithuania (grant No.~VIZIT-2-TYR-005).

The research of K. Pileckas was funded by a grant No.~MIP-030/2011
from the Research Council of Lithuania.

The research of R. Russo was supported by "Gruppo Nazionale per la
Fisica Matematica" of  "Istituto Nazionale di Alta Matematica".


{\small
\begin{thebibliography}{99}

\bibitem{Amick} {\sc Ch.J. Amick}: Existence of solutions to the
nonhomogeneous steady Navier--Stokes equations, {\it Indiana Univ.
Math. J.\/} {\bf 33} (1984), 817--830.


\bibitem{BOPI}{\sc W. Borchers and K. Pileckas}: Note on the flux
problem for stationary Navier--Stokes equations in domains with
multiply connected boundary, {\it Acta App. Math.\/} {\bf 37}
(1994), 21--30.

\bibitem{korob}{\sc J. Bourgain, M.V. Korobkov and J. Kristensen}:
On the Morse-- Sard property and level sets of Sobolev and BV
functions, {\it arXiv:1007.4408v1\/}, [math.AP], 26 July 2010.

\bibitem{CLMS}
{\sc R.R. Coifman, J.L. Lions, Y. Meier and S. Semmes}:
Compensated compactness and Hardy spaces, {\sl J. Math. Pures
App.\/} IX S\'er. 72  (1993), 247--286.

\bibitem{Dor} {\sc J. R. Dorronsoro}: Differentiability properties of functions with bounded
variation, {\it Indiana U. Math. J.\/} {\bf 38}, no.~4 (1989),
1027--1045.

\bibitem{evans} {\sc L.C. Evans, R.F. Gariepy}:
{\it Measure theory and fine properties of functions}, Studies in Advanced
Mathematics. CRC Press, Boca Raton, FL  (1992).

\bibitem{Finn}{\sc R. Finn}: On the steady-state solutions of the
Navier--Stokes equations. III, {\it Acta Math.} {\bf 105} (1961),
197--244.

\bibitem{Fu}{\sc H. Fujita}: On the existence and regularity of the
steady-state solutions of the Navier--Stokes theorem, {\it J. Fac.
Sci. Univ. Tokyo Sect.\/} I (1961) {\bf 9}, 59--102.

\bibitem{Fu1} {\sc H. Fujita}: On stationary solutions to Navier--Stokes equation in
symmetric plane domain under general outflow condition, {\it
Pitman research notes in mathematics, Proceedings of International
conference on Navier--Stokes equations. Theory and numerical
methods. June 1997. Varenna, Italy\/} (1997) {\bf 388}, 16--30.

\bibitem{Galdi1}{\sc  G.P. Galdi}: On the existence of steady
motions of a viscous flow with non--homogeneous conditions, {\it Le
Matematiche\/} {\bf 66} (1991), 503--524.

\bibitem{Galdibook}{\sc  G.P. Galdi}: {\it An introduction to the mathematical theory of the Navier-
Stokes equation. Steady-state problems}, second edition, Springer
(2011).

\bibitem{KaPi1} {\sc  L.V. Kapitanskii and K. Pileckas}: On spaces
of solenoidal vector fields and boundary value problems for the
Navier--Stokes equations in domains with noncompact boundaries,
{\it Trudy Mat. Inst. Steklov \/} {\bf 159}  (1983), 5--36 .
English Transl.: {\it Proc. Math. Inst. Steklov} {\bf 159}
(1984),   3--34.

\bibitem{korob1} {\sc  M.V. Korobkov}: Bernoulli law under
minimal smoothness assumptions, {\it Dokl. Math. \/} {\bf 83}
(2011), 107--110.

\bibitem{kpr} {\sc M.V. Korobkov, K. Pileckas and R. Russo},  On the flux problem in the theory of steady
Navier--Stokes equations with nonhomogeneous boundary conditions,
arXiv:1009.4024

\bibitem{kpr_a_crm} {\sc M.V. Korobkov, K. Pileckas and R. Russo},  Steady Navier-Stokes system with nonhomogeneous boundary conditions in the
axially symmetric case, {\it Comptes rendus~-- Mecanique
 \/} {\bf 340} (2012), 115--119.


 \bibitem{kpr_a_arx} {\sc M.V. Korobkov, K. Pileckas and R. Russo},  Steady Navier-Stokes system with nonhomogeneous boundary conditions in the
axially symmetric case, arXiv:1110.6301.

\bibitem{Kozono}{\sc H. Kozono and T. Yanagisawa}: Leray's problem on
the stationary Navier--Stokes equations with inhomogeneous
boundary data, {\sl Math. Z. \/} {\bf 262} No. 1 (2009), 27--39 .

\bibitem{Kronrod} {\sc  A.S. Kronrod}: On functions of two
variables, {\sl Uspechi Matem. Nauk (N.S.)\/} {\bf 5} (1950),
24--134 (in Russian).

\bibitem{Lad1} {\sc O.A. Ladyzhenskaya}: Investigation of the
Navier--Stokes equations in the case of stationary motion of an
incompressible fluid, {\it Uspech Mat. Nauk\/}  {\bf 3} (1959),
75--97 (in Russian).

\bibitem{Lad} {\sc O.A. Ladyzhenskaya}: {\it The Mathematical theory
of viscous  incompressible fluid}, Gordon and Breach (1969).

\bibitem{LadSol1} {\sc O.A. Ladyzhenskaya and V.A. Solonnikov}:
On some problems of vector analysis and generalized formulations
of boundary value problems for the Navier--Stokes equations, {\sl
Zapiski Nauchn. Sem. LOMI \/} {\bf 59} (1976), 81--116 (in
Russian); English translation in {\sl Journal of Soviet
Mathematics \/}  {\bf 10} (1978), no.2, 257--286.

\bibitem{Leray} {\sc J. Leray}:  \'Etude de diverses
\'equations int\'egrales non lin\'eaire  et de quelques
probl\`emes que pose l'hydrodynamique, {\it J. Math. Pures
Appl.\/} {\bf 12} (1933), 1--82.

\bibitem{maz'ya} {\sc V.G. Maz'ya}: {\sl Sobolev Spaces}, Springer-Verlag
(1985).

\bibitem{Mi2}{\sc C. Miranda}: {\it Partial differential equations of elliptic
type}, Springer--Verlag (1970).

\bibitem{Morimoto}{\sc H. Morimoto}: A remark on the existence of 2--D steady Navier--Stokes
flow in bounded symmetric domain under general outflow condition,
{\it J. Math. Fluid Mech.\/} {\bf 9}, No. 3 (2007), 411--418.

\bibitem{Neustupa}
{\sc J. Neustupa}: A new approach to the existence of weak
solutions of the steady Navier--Stokes system with inhomoheneous
boundary data in domains with noncompact boundaries, {\sl Arch.
Rational Mech. Anal\/} {\bf 198}, No. 1 (2010), 331--348.


\bibitem{Pukhnachev} {\sc  V.V. Pukhnachev}: Viscous flows in domains with a multiply connected
boundary, {\it New Directions in Mathematical Fluid Mechanics. The
Alexander V. Kazhikhov Memorial Volume. Eds. Fursikov A.V., Galdi
G.P. and Pukhnachev V.V. Basel -- Boston -- Berlin: Birkhauser\/}
(2009) 333--348.

\bibitem{Pukhnachev1} {\sc  V.V. Pukhnachev}: The Leray problem
and the Yudovich hypothesis, {\it Izv. vuzov. Sev.--Kavk. region.
Natural sciences. The special issue "Actual problems of
mathematical hydrodynamics"\/} (2009) 185--194 (in Russian).

\bibitem{RussoA}{\sc A. Russo}: A note on  the  two--dimensional steady-state
Navier--Stokes problem, {\sl J. Math. Fluid Mech.\/}, {\bf 11}
(2009) 407--414.

\bibitem{Russo}{\sc R. Russo}: On the existence of solutions  to the stationary
Navier--Stokes equations, {\it Ricerche Mat.\/} {\bf 52} (2003),
285--348.

\bibitem{SolSca}
{\sc V. A. Solonnikov  and V. E. Scadilov}: On a boundary value
problem for a stationary system of Navier--Stokes equations, {\it
Proc. Steklov Inst. Math.\/} {\bf 125} (1973), 186--199 (in
Russian).

\bibitem{Stein}
{\sc  E. Stein}: {\sl Harmonic analysis: real--variables methods,
orthogonality and oscillatory integrals\/}, Princeton University
Press (1993).

\bibitem{Takashita} {\sc A. Takashita}: A remark on Leray's
inequality,  {\it Pacific J. Math.\/} {\bf 157} (1993), 151--158.

\bibitem{VorJud} {\sc I.I. Vorovich and V.I. Yudovich}: Stationary
flows of a viscous incompres-sible fluid, {\it Mat. Sbornik\/} {\bf
53} (1961), 393--428 (in Russian).

\end{thebibliography} }
\end{document}